\documentclass[aps,twocolumn,showpacs,superscriptaddress,groupedaddress,amsmath]{revtex4}

\usepackage{graphicx}
\usepackage[usenames]{color}
\usepackage{amssymb,amsmath}

\newcommand{\pt}        {\mbox{$p_T$}}
\newcommand{\met}       {\mbox{$\not\!\!E_T$}}

\newcommand{\rargap}    {\mbox{ $\rightarrow$ }}
\newcommand{\ppbar}     {\mbox{$p\bar{p}$}}
\newcommand{\ttbar}     {\mbox{$t\bar{t}$}}
\newcommand{\bbbar}     {\mbox{$b\bar{b}$}}
\newcommand{\ccbar}     {\mbox{$c\bar{c}$}}
\newcommand{\comphep}   {\sc comphep}
\newcommand{\singletop} {\sc singletop}
\newcommand{\pythia}    {\sc pythia}
\newcommand{\alpgen}    {\sc alpgen}
\newcommand{\mcfm}      {\sc mcfm}
\newcommand{\geant}     {\sc geant}

\begin{document}

\title{Measurements of single top quark production cross sections and 
$\boldsymbol{|V_{tb}|}$ in $\boldsymbol{p\bar{p}}$ collisions at $\boldsymbol{\sqrt{s}=1.96}$~TeV }

\affiliation{Universidad de Buenos Aires, Buenos Aires, Argentina}
\affiliation{LAFEX, Centro Brasileiro de Pesquisas F{\'\i}sicas, Rio de Janeiro, Brazil}
\affiliation{Universidade do Estado do Rio de Janeiro, Rio de Janeiro, Brazil}
\affiliation{Universidade Federal do ABC, Santo Andr\'e, Brazil}
\affiliation{Instituto de F\'{\i}sica Te\'orica, Universidade Estadual Paulista, S\~ao Paulo, Brazil}
\affiliation{University of Science and Technology of China, Hefei, People's Republic of China}
\affiliation{Universidad de los Andes, Bogot\'{a}, Colombia}
\affiliation{Charles University, Faculty of Mathematics and Physics, Center for Particle Physics, Prague, Czech Republic}
\affiliation{Czech Technical University in Prague, Prague, Czech Republic}
\affiliation{Center for Particle Physics, Institute of Physics, Academy of Sciences of the Czech Republic, Prague, Czech Republic}
\affiliation{Universidad San Francisco de Quito, Quito, Ecuador}
\affiliation{LPC, Universit\'e Blaise Pascal, CNRS/IN2P3, Clermont, France}
\affiliation{LPSC, Universit\'e Joseph Fourier Grenoble 1, CNRS/IN2P3, Institut National Polytechnique de Grenoble, Grenoble, France}
\affiliation{CPPM, Aix-Marseille Universit\'e, CNRS/IN2P3, Marseille, France}
\affiliation{LAL, Universit\'e Paris-Sud, CNRS/IN2P3, Orsay, France}
\affiliation{LPNHE, Universit\'es Paris VI and VII, CNRS/IN2P3, Paris, France}
\affiliation{CEA, Irfu, SPP, Saclay, France}
\affiliation{IPHC, Universit\'e de Strasbourg, CNRS/IN2P3, Strasbourg, France}
\affiliation{IPNL, Universit\'e Lyon 1, CNRS/IN2P3, Villeurbanne, France and Universit\'e de Lyon, Lyon, France}
\affiliation{III. Physikalisches Institut A, RWTH Aachen University, Aachen, Germany}
\affiliation{Physikalisches Institut, Universit{\"a}t Freiburg, Freiburg, Germany}
\affiliation{II. Physikalisches Institut, Georg-August-Universit{\"a}t G\"ottingen, G\"ottingen, Germany}
\affiliation{Institut f{\"u}r Physik, Universit{\"a}t Mainz, Mainz, Germany}
\affiliation{Ludwig-Maximilians-Universit{\"a}t M{\"u}nchen, M{\"u}nchen, Germany}
\affiliation{Fachbereich Physik, Bergische Universit{\"a}t Wuppertal, Wuppertal, Germany}
\affiliation{Panjab University, Chandigarh, India}
\affiliation{Delhi University, Delhi, India}
\affiliation{Tata Institute of Fundamental Research, Mumbai, India}
\affiliation{University College Dublin, Dublin, Ireland}
\affiliation{Korea Detector Laboratory, Korea University, Seoul, Korea}
\affiliation{CINVESTAV, Mexico City, Mexico}
\affiliation{Nikhef, Science Park, Amsterdam, the Netherlands}
\affiliation{Radboud University Nijmegen, Nijmegen, the Netherlands and Nikhef, Science Park, Amsterdam, the Netherlands}
\affiliation{Joint Institute for Nuclear Research, Dubna, Russia}
\affiliation{Institute for Theoretical and Experimental Physics, Moscow, Russia}
\affiliation{Moscow State University, Moscow, Russia}
\affiliation{Institute for High Energy Physics, Protvino, Russia}
\affiliation{Petersburg Nuclear Physics Institute, St. Petersburg, Russia}
\affiliation{Instituci\'{o} Catalana de Recerca i Estudis Avan\c{c}ats (ICREA) and Institut de F\'{i}sica d'Altes Energies (IFAE), Barcelona, Spain}
\affiliation{Stockholm University, Stockholm and Uppsala University, Uppsala, Sweden}
\affiliation{Lancaster University, Lancaster LA1 4YB, United Kingdom}
\affiliation{Imperial College London, London SW7 2AZ, United Kingdom}
\affiliation{The University of Manchester, Manchester M13 9PL, United Kingdom}
\affiliation{University of Arizona, Tucson, Arizona 85721, USA}
\affiliation{University of California Riverside, Riverside, California 92521, USA}
\affiliation{Florida State University, Tallahassee, Florida 32306, USA}
\affiliation{Fermi National Accelerator Laboratory, Batavia, Illinois 60510, USA}
\affiliation{University of Illinois at Chicago, Chicago, Illinois 60607, USA}
\affiliation{Northern Illinois University, DeKalb, Illinois 60115, USA}
\affiliation{Northwestern University, Evanston, Illinois 60208, USA}
\affiliation{Indiana University, Bloomington, Indiana 47405, USA}
\affiliation{Purdue University Calumet, Hammond, Indiana 46323, USA}
\affiliation{University of Notre Dame, Notre Dame, Indiana 46556, USA}
\affiliation{Iowa State University, Ames, Iowa 50011, USA}
\affiliation{University of Kansas, Lawrence, Kansas 66045, USA}
\affiliation{Kansas State University, Manhattan, Kansas 66506, USA}
\affiliation{Louisiana Tech University, Ruston, Louisiana 71272, USA}
\affiliation{Boston University, Boston, Massachusetts 02215, USA}
\affiliation{Northeastern University, Boston, Massachusetts 02115, USA}
\affiliation{University of Michigan, Ann Arbor, Michigan 48109, USA}
\affiliation{Michigan State University, East Lansing, Michigan 48824, USA}
\affiliation{University of Mississippi, University, Mississippi 38677, USA}
\affiliation{University of Nebraska, Lincoln, Nebraska 68588, USA}
\affiliation{Rutgers University, Piscataway, New Jersey 08855, USA}
\affiliation{Princeton University, Princeton, New Jersey 08544, USA}
\affiliation{State University of New York, Buffalo, New York 14260, USA}
\affiliation{Columbia University, New York, New York 10027, USA}
\affiliation{University of Rochester, Rochester, New York 14627, USA}
\affiliation{State University of New York, Stony Brook, New York 11794, USA}
\affiliation{Brookhaven National Laboratory, Upton, New York 11973, USA}
\affiliation{Langston University, Langston, Oklahoma 73050, USA}
\affiliation{University of Oklahoma, Norman, Oklahoma 73019, USA}
\affiliation{Oklahoma State University, Stillwater, Oklahoma 74078, USA}
\affiliation{Brown University, Providence, Rhode Island 02912, USA}
\affiliation{University of Texas, Arlington, Texas 76019, USA}
\affiliation{Southern Methodist University, Dallas, Texas 75275, USA}
\affiliation{Rice University, Houston, Texas 77005, USA}
\affiliation{University of Virginia, Charlottesville, Virginia 22901, USA}
\affiliation{University of Washington, Seattle, Washington 98195, USA}
\author{V.M.~Abazov} \affiliation{Joint Institute for Nuclear Research, Dubna, Russia}
\author{B.~Abbott} \affiliation{University of Oklahoma, Norman, Oklahoma 73019, USA}
\author{B.S.~Acharya} \affiliation{Tata Institute of Fundamental Research, Mumbai, India}
\author{M.~Adams} \affiliation{University of Illinois at Chicago, Chicago, Illinois 60607, USA}
\author{T.~Adams} \affiliation{Florida State University, Tallahassee, Florida 32306, USA}
\author{G.D.~Alexeev} \affiliation{Joint Institute for Nuclear Research, Dubna, Russia}
\author{G.~Alkhazov} \affiliation{Petersburg Nuclear Physics Institute, St. Petersburg, Russia}
\author{A.~Alton$^{a}$} \affiliation{University of Michigan, Ann Arbor, Michigan 48109, USA}
\author{G.~Alverson} \affiliation{Northeastern University, Boston, Massachusetts 02115, USA}
\author{G.A.~Alves} \affiliation{LAFEX, Centro Brasileiro de Pesquisas F{\'\i}sicas, Rio de Janeiro, Brazil}
% ...
%
% visitor_addresses.tex                        2 August 2011
%  available symbols are:
%  $\ast, \dag, \ddag, \S, \P, $\|$, $\ast\ast$, \dag\dag, \ddag\ddag ,\#
%
\collaboration{The D0 Collaboration\footnote{with visitors from
%{alton}
$^{a}$Augustana College, Sioux Falls, SD, USA,
%{burdin}
$^{b}$The University of Liverpool, Liverpool, UK,
%{falkowski}
%$^{?}$Laboratoire de Physique Theorique, Orsay, FR
%{garcia-guerra}
$^{c}$UPIITA-IPN, Mexico City, Mexico,
%{haas,partridge}
$^{c}$SLAC, Menlo Park, CA, USA,
%{hesketh}
$^{e}$University College London, London, UK,
%{luna-garcia}
$^{f}$Centro de Investigacion en Computacion - IPN, Mexico City, Mexico,
%{podesta-lerma}
$^{g}$ECFM, Universidad Autonoma de Sinaloa, Culiac\'an, Mexico,
and 
%{weber}
$^{h}$Universit{\"a}t Bern, Bern, Switzerland.
%{hooper}
%$^{?}$Visitor from Bradley University, Peoria, IL, USA.
%{kozminski}
%$^{?}$}Visitor from Lewis University, Romeoville, IL, USA.
%{deceased}
$^{\ddag}$Deceased.
}} \noaffiliation
\vskip 0.25cm

\date{August 15, 2011}

\begin{abstract}
We present measurements of production cross sections of single top quarks in $\ppbar$ collisions 
at $\sqrt{s}=1.96\;\rm TeV$ in a data sample corresponding to an integrated luminosity of 
$5.4\;\rm fb^{-1}$ collected by the D0 detector at the Fermilab Tevatron Collider. 
We select events with an isolated electron or muon, an imbalance in transverse energy, 
and two, three, or four jets, with one or two of them containing a bottom hadron. We obtain an inclusive cross section 
of $\sigma({\ppbar}{\rargap}tb+X,~tqb+X) = 3.43\pm^{0.73}_{0.74}\;\rm pb$ and use it to 
extract the CKM matrix element $0.79 < |V_{tb}| \leq 1$ at the 95\% C.L. We also measure 
$\sigma({\ppbar}{\rargap}tb+X) = 0.68\pm^{0.38}_{0.35}\;\rm pb$ and 
$\sigma({\ppbar}{\rargap}tqb+X) = 2.86\pm^{0.69}_{0.63}\;\rm pb$ 
when assuming, respectively, $tqb$ and $tb$ production rates as predicted by the standard model.
\end{abstract}


\maketitle 

\section{Introduction}

Top quarks are produced at hadron colliders as $\ttbar$ pairs via the strong interaction or singly via the electroweak 
interaction~\cite{ttbar-xsec,singletop-xsec-kidonakis}. Because of the larger production rate and higher 
signal-to-background ratio, the production of $\ttbar$ pairs is better studied and indeed  
it was through the $\ttbar$ production process
that the existence of the top quark was established in 1995 at the Fermilab 
Tevatron Collider~\cite{top-obs-1995-d0,top-obs-1995-cdf}. 
The observation of the single top quark production, however, was possible 
after CDF and D0 collaborations accumulated $\approx50$ times more integrated luminosity than what was needed for
observation of top quarks in $\ttbar$ production~\cite{stop-obs-2009-cdf,stop-obs-2009-d0}.
Single top quark events are produced at about half of the rate of top quark pairs and 
with lower jet multiplicities, and therefore their study is more susceptible to contamination from background processes.

\begin{figure}
\vspace{0.1in}
\includegraphics[width=0.50\textwidth]{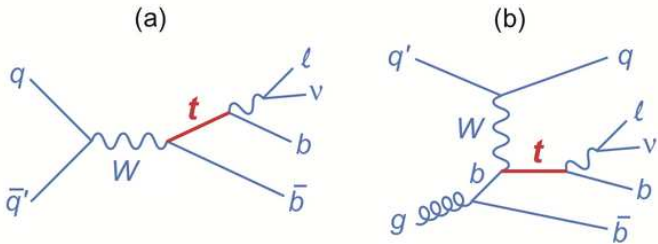}
\vspace{-0.1in}
\caption{[color online] Lowest level Feynman diagrams for (a) $tb$ and (b) $tqb$ single top quark production.}
\label{fig:feynman}
\end{figure}

Electroweak production of top quarks at the Tevatron proceeds mainly via the decay of a time-like virtual $W$ boson 
accompanied by a bottom quark in the $s$ channel ($tb = t\bar{b}+\bar{t}b$)~\cite{singletop-cortese} 
or via the exchange of a space-like virtual $W$ boson between a light quark and a bottom quark in the 
$t$ channel ($tqb = tq\bar{b}+\bar{t}qb$, where $q$ refers to the light quark or antiquark)~\cite{singletop-willenbrock,singletop-yuan}.
Figure~\ref{fig:feynman} shows the lowest level Feynman diagrams for $s$- and $t$-channel production~\cite{charge-conjugate}. 
A third process $tW$, in which the top quark is produced together 
with a $W$ boson, has a small cross section at the Tevatron~\cite{singletop-xsec-kidonakis} and is therefore 
not considered in this analysis.

Single top quark events can be used to probe the $Wtb$ vertex and to directly measure 
the magnitude of the Cabibbo-Kobayashi-Maskawa (CKM)~\cite{CKM} quark mixing matrix element $|V_{tb}|$.
Under the assumptions that there are only three quark generations and
that the CKM matrix is unitary, the matrix elements 
are severely constrained~\cite{pdg}: $|V_{td}|=(8.62^{+0.26}_{-0.20}) \times 10^{-3}$, $|V_{ts}|=(4.03^{+0.11}_{-0.07}) \times 10^{-4}$, 
and $|V_{tb}|=0.999152^{+0.000030}_{-0.000045}$. However, in several extensions of the standard model (SM) involving, for 
instance, a fourth generation of quarks or an additional heavy quark singlet that mixes with the top quark, 
$|V_{tb}|$ can be significantly smaller than unity~\cite{Alwall:2007}. A direct determination of $|V_{tb}|$,
without assuming unitarity or three generations, is possible through the measurement of the total single top quark production cross 
section~\cite{singletop-vtb-jikia}. The current measured 
value for the total single top quark cross section 
is $2.76^{+0.58}_{-0.47}\;\rm pb$ at $\sqrt{s}=1.96\;\rm TeV$, resulting in $|V_{tb}| = 0.88 \pm 0.07$ 
with a limit of $|V_{tb}| > 0.77$ at the 95\% C.L. 
assuming a top quark mass $m_t =  170$~GeV~\cite{TeVComb}.
 
Previous measurements of single top quark production cross 
sections~\cite{d0-prl-2007, d0-prd-2008, stop-obs-2009-d0, cdf-prl-2008, stop-obs-2009-cdf, stop-obs-2009-long-cdf} 
included events from both the $tb$ and $tqb$ processes, assuming a ratio of 
cross sections~\cite{singletop-xsec-kidonakis} for the two processes based on the SM.
However, several beyond-the-SM theories predict individual $tb$ and $tqb$ cross sections that deviate from the SM.
Examples include models with additional quark generations~\cite{Alwall:2007}, new heavy bosons~\cite{Tait:2000sh}, 
flavor-changing neutral currents (FCNC)~\cite{d0-fcnc}, or anomalous top quark 
couplings~\cite{singletop-wtb-heinson,d0-singletop-wtb,Abazov:2009ky}. 
It is therefore important to also measure the individual 
$tb$ and $tqb$ production rates. 
 
Using data corresponding to $5.4\;\rm fb^{-1}$ of integrated 
luminosity recorded with the D0 detector~\cite{d0-detector}, we 
present an improved measurement of the production rate of $tb$+$tqb$. We also present 
measurements of the production rates of the individual $tb$ and $tqb$ processes performed assuming,  
respectively, $tqb$ and $tb$ production rates as predicted by the SM.
Finally, we present a new direct measurement of $|V_{tb}|$ extracted 
from the measured $tb$+$tqb$ cross section. 

\section{Event selection}
\label{sec:event-selection-cuts}

This analysis extends previous work by the D0 
Collaboration~\cite{d0-prl-2007, d0-prd-2008, stop-obs-2009-d0, t-channel} and uses the same 
data, event selection, and modeling of signal and background as in Ref.~\cite{t-channel-new}. It differs 
however, in the assumptions used to extract the cross sections of the individual $tb$ and $tqb$ production modes. 

The data were collected with a logical OR of many trigger conditions, which together are fully efficient 
for the single top quark signal. We select events containing only one isolated 
electron or muon with high transverse momentum ($\pt$) and having a large imbalance
in the transverse energy ($\met$) indicative of the presence of a 
neutrino. Events originating from single top quark production are expected to contain at least one 
$b$ quark jet from the decay of the top quark and a second $b$ quark jet in the $s$ channel, or a light 
quark jet and a spectator $b$ quark jet for the $t$ channel. In both cases, gluon radiation can give rise to additional 
jets. Events are selected with two, three, or four jets reconstructed using a cone algorithm~\cite{Cone}  
in $(y,\phi)$ space, where $y$ is the rapidity and $\phi$ is the azimuthal angle, and the cone radius $0.5$.
The jets must satisfy the following conditions: leading jet $\pt > 25\;\rm GeV$, other jets with $\pt > 15\;\rm GeV$, 
and with  pseudorapidities of all jets $|\eta| < 3.4$. Requirements are also placed on $\met$: 
$20 < \met < 200\;\rm GeV$ for events with two jets, and $25 < \met < 200\;\rm GeV$ for events with three 
or four jets. The maximum $\met$ requirement removes events that suffer from poor modeling of the high energy tail 
of the muon momentum resolution.
We require one isolated electron with $|\eta| < 1.1$ and $\pt > 15\;\rm (20)\;\rm GeV$ for events with
 two (three or four) jets, or one isolated muon with $|\eta| < 2.0$ and $\pt > 15\;\rm GeV$. 
 
The sample resulting from this selection is dominated by $W$ bosons produced in association with 
jets ($W$+jets), with smaller contributions from {\ttbar} pairs decaying into the single lepton plus jets final state or 
the dilepton final state when one lepton or some jets are not reconstructed. Multijet events also
contribute to the background when a jet is misidentified as an isolated electron or a heavy-flavor 
quark decays to a muon that satisfies isolation criteria, in combination with misreconstruction of $\met$.
Diboson ($WW$, $WZ$, $ZZ$) and $Z$+jets processes 
contribute only marginally to the total background but are taken into account. 
The background from multijets is kept small ($< 6$\%) by requiring that the total 
scalar sum ($H_T$) of the transverse momenta of the final-state objects (lepton, $\met$, and jets) 
be larger than 120/140/160~GeV for events with 2/3/4 jets, and that the $\met$ 
does not point along the transverse direction of the lepton or the leading jet. 
Soft-scattering processes are suppressed by requiring a minimum value for the total scalar sum of the $\pt$ of the jets
[$H_T ({\rm alljets})$] ranging from 50 to 100~GeV, depending on the number of jets in an event and the data collection period. 

To enhance the signal fraction, one or two of the jets are required to contain
long lived bottom hadrons ($b$ jets), as determined through a multivariate $b$-tagging algorithm~\cite{D0btag}. 
This algorithm uses several variables to discriminate $b$ jets from other jets such as:
(i) decay length significance of the secondary vertex, 
(ii) the $\chi^2$ per degree of freedom of the secondary vertex fit,
(iii) weighted combination of the tracks' impact parameter significances,
(iv) probability that the jet originates from the primary {\ppbar} interaction vertex, 
(v) number of tracks used to reconstruct the secondary vertex, 
(vi) mass of the secondary vertex, and 
(vii) number of secondary vertices found inside the jet.
To improve sensitivity to signal, the samples are divided into six independent analysis channels, depending on the 
jet multiplicity (two, three, or four jets), and the number of $b$-tagged jets (one or two). 
The efficiency of the event selection, including branching fraction and the $b$-tagging requirements, is $(2.9 \pm 0.4)\%$ for $tb$ 
and $(2.0 \pm 0.3)\%$ for $tqb$. 
The $tqb$ process has a lower acceptance than the $tb$ channel 
because the second $b$-jet has low transverse momentum and is difficult to identify. 
We apply additional requirements to select two control samples used to test whether the background model
reproduces the data in regions dominated by one specific type of background. 
The control sample dominated by $W$+jets is required to have exactly two jets, $H_T < 175\;\rm GeV$, and 
only one $b$-tagged jet where $W$+jets events constitute 82\% of this sample, and the {\ttbar} component is less that
2\%. The control sample dominated by {\ttbar} is required to have exactly four jets, 
$H_T > 300\;\rm GeV$, and one or two $b$-tagged jets where {\ttbar} events constitute 84\% of the sample, and the 
$W$+jets component is 12\%.

\section{Models for signal and background}

Single top quark events are modeled for a top quark mass $m_t =  172.5$~GeV
using the {\comphep}-based effective next-to-leading order (NLO) Monte Carlo (MC) 
event generator {\singletop}~\cite{singletop-mcgen}, which preserves spin information in the decays 
of the top quark and the $W$ boson and provides event kinematics that reproduce distributions 
predicted by NLO calculations~\cite{singletop-xsec-sullivan,Campbell:2009ss}.  
The {\ttbar}, $W$+jets, and $Z$+jets events are simulated with the {\alpgen} leading orden MC 
generator~\cite{alpgen}. Diboson processes are modeled using {\pythia}~\cite{pythia}. For all these MC 
samples, {\pythia} is also used to evolve parton showers and to model proton remnants and hadronization of all generated 
partons. 
The presence of additional $\ppbar$ interactions is modeled
by events selected from random beam crossings matching
the instantaneous luminosity profile in the data.
All MC events are passed through a {\geant}-based simulation~\cite{geant} of the D0 detector. 

Differences between simulation and data in lepton and jet reconstruction efficiencies and resolutions, 
jet energy scale, and $b$-tagging efficiencies are corrected in the simulation by applying correction functions 
measured from separate data samples. 
Comparisons of {\alpgen} with data and with other generators show 
small discrepancies in distributions of jet pseudorapidity
and angular separations between jets~\cite{reweight}.
We therefore correct the {\alpgen} $W$+jets and $Z$+jets samples by sequentially applying 
polynomial reweighting functions parameterized
by the leading and second-leading jet $\eta$, $\Delta R = \sqrt{(\Delta\phi)^2+(\Delta\eta)^2}$ 
between the two leading jets, and third- and fourth-leading jet $\eta$, if applicable.
These functions are derived from the ratio between the number of $W$+jets and $Z$+jets events
observed in data and the event yields predicted by MC.
After these corrections, the MC description is in good agreement with our high 
statistics sample of events prior to the application of $b$-tagging. 
The multijet background is modeled using the selection discussed in Sec.~\ref{sec:event-selection-cuts}, 
but choosing events that fail isolation criteria for leptons.

MC samples are scaled to the theoretical cross section at 
approximately NNLO~\cite{ttbar-xsec} for {\ttbar}, and NLO~\cite{mcfm} for $Z$+jets and diboson cases.   
The contributions from $W$+jets and multijet are normalized by comparing the prediction for
background to data before $b$-tagging. We use a procedure that relies on three distributions 
[lepton $p_{T}$, $\met$ and $W$ reconstructed mass in the transverse plane $M_T(W)$]
that have distinctive shapes for $W$+jets and multijets events 
and are thus sensitive to their relative contributions in the selected sample.
The normalization scale factors for $W$+jets ($\lambda_{\text{Wjets}}$) and 
multijet ($\lambda_{\text{multijets}}$) are constrained by the following equation:
\begin{equation}
N = \lambda_{\text{Wjets}} N_{\text{Wjets}} + \lambda_{\text{multijets}} N_{\text{multijets}},
\label{eq:wjetsqcd}
\end{equation}
where $N = N_{\text{data}} - N_{\text{non-Wjets}}$ and $N_{\text{data}}$, $N_{\text{non-Wjets}}$, 
$N_{\text{Wjets}}$, and $N_{\text{multijets}}$ are the event yields in data, 
non-$W$+jets MC, $W$+jets, and multijet samples, respectively.
The $W$+jets sample contains events with light flavor ($Wjj$, $j=u,d,s$) 
and heavy flavor ($Wjc$, $Wc\bar{c}$, and $Wb\bar{b}$) quarks.       
The non-$W$+jets MC samples include single top quark, {\ttbar}, $Z$+jets, and diboson production. 
The values of $\lambda_{\text{Wjets}}$ and $\lambda_{\text{multijets}}$ are
varied to maximize the product of the Kolmogorov-Smirnov test
values~\cite{KStest} for the three kinematic distributions.
This procedure is done separately for events with two, 
three, and four jets and for each lepton flavor.
After the normalization, the total sum of the $W$+jets and multijets yields plus the small 
contributions from {\ttbar}, single top, $Z$+jets, and diboson production equals the total data yield for each of the six analysis channels. 

Without modifying the overall normalization of the $W$+jets MC sample, we 
apply an additional scale factor to $W$ and $Z$ boson events produced in conjunction 
with heavy-flavor jets ($b$ or $c$) to match NLO calculations~\cite{mcfm}: 
$W\bbbar$ and $W\ccbar$ by 1.47, $Z\bbbar$ by 1.52, $Z\ccbar$ by 1.67, and $Wcj$ by 1.32.
We evaluate whether an additional normalization factor $\lambda_{HF}$ is required 
for the $W\bbbar$ and $W\ccbar$ samples by using events with two jets that pass the event selection 
described in Sec.~\ref{sec:event-selection-cuts} but fail the $b$-tagging requirements (zero-tag sample). 
The zero-tag sample has no overlap with the sample used to measure the single top quark cross section. 
During this study, we 
keep the normalization of the $W+$jets sample fixed to the value obtained by the iterative method described
above and derive $\lambda_{\text{HF}}$ with the following equation: 
\begin{equation}
\label{eq:CombinedHFCountingZeroTag_Equation}
N^{(0)} = N^{(0)}_{\text{Wlp}} + \lambda_{\text{HF}} N^{(0)}_{\text{Whp}},
\end{equation}
where $N = N_{\text{data}} - N_{\text{multijets}} - N_{\text{non-Wjets}}$, 
$N_{\text{Wlp}} = N_{\text{Wjj}} + N_{\text{Wcj}}$, and 
$N_{\text{Whp}} = N_{\text{Wcc}} + N_{\text{Wbb}}$. 
The superscript $(0)$ indicates that the equation is written for the zero-tag sample defined above.
The measured value of $\lambda_{\text{HF}}$ is consistent with one. 
Uncertainties on the assumed cross sections for single top quark, \ttbar, and $Wcj$ production and the cross section ratio of 
$Wc\bar{c}$ to $Wb\bar{b}$ are taken into account. 
As expected, $\lambda_{HF}$ is most affected by variations on the $Wcj$ cross section and the 
$Wc\bar{c}$ to $Wb\bar{b}$ cross section ratio. An estimated uncertainty of 12\% is assigned to the normalization of the $Wc\bar{c}$ and $Wb\bar{b}$ MC samples based on this study.

We also consider other sources of systematic uncertainty from modeling both the background and signal. These uncertainties usually affect the 
normalization and, in some cases, also the shape of the distributions. The largest uncertainties 
arise from the jet energy scale (0.3--14.6)\%,  jet energy resolution (0.2--11.6)\%, corrections to $b$-tagging efficiencies (6.6--21.2)\%, and the correction for jet-flavor composition in $W$+jets events 12\%. There are also contributions due to limited statistics of the MC samples 6.0\%, the measured luminosity 6.1\%, and uncertainties on the trigger modeling 5.0\%.  

\begin{figure}
\centering
\includegraphics[width=0.235\textwidth]{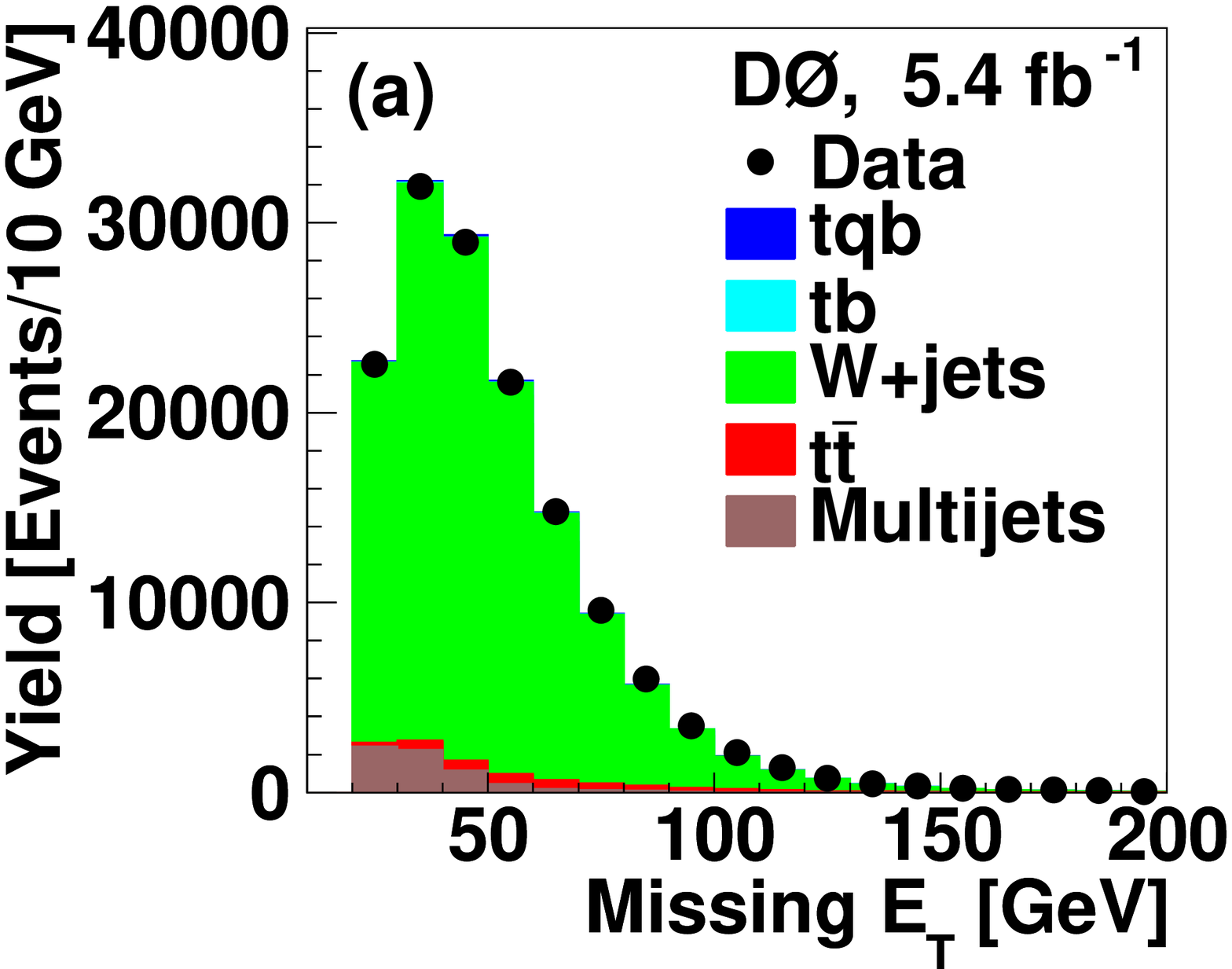}
\includegraphics[width=0.235\textwidth]{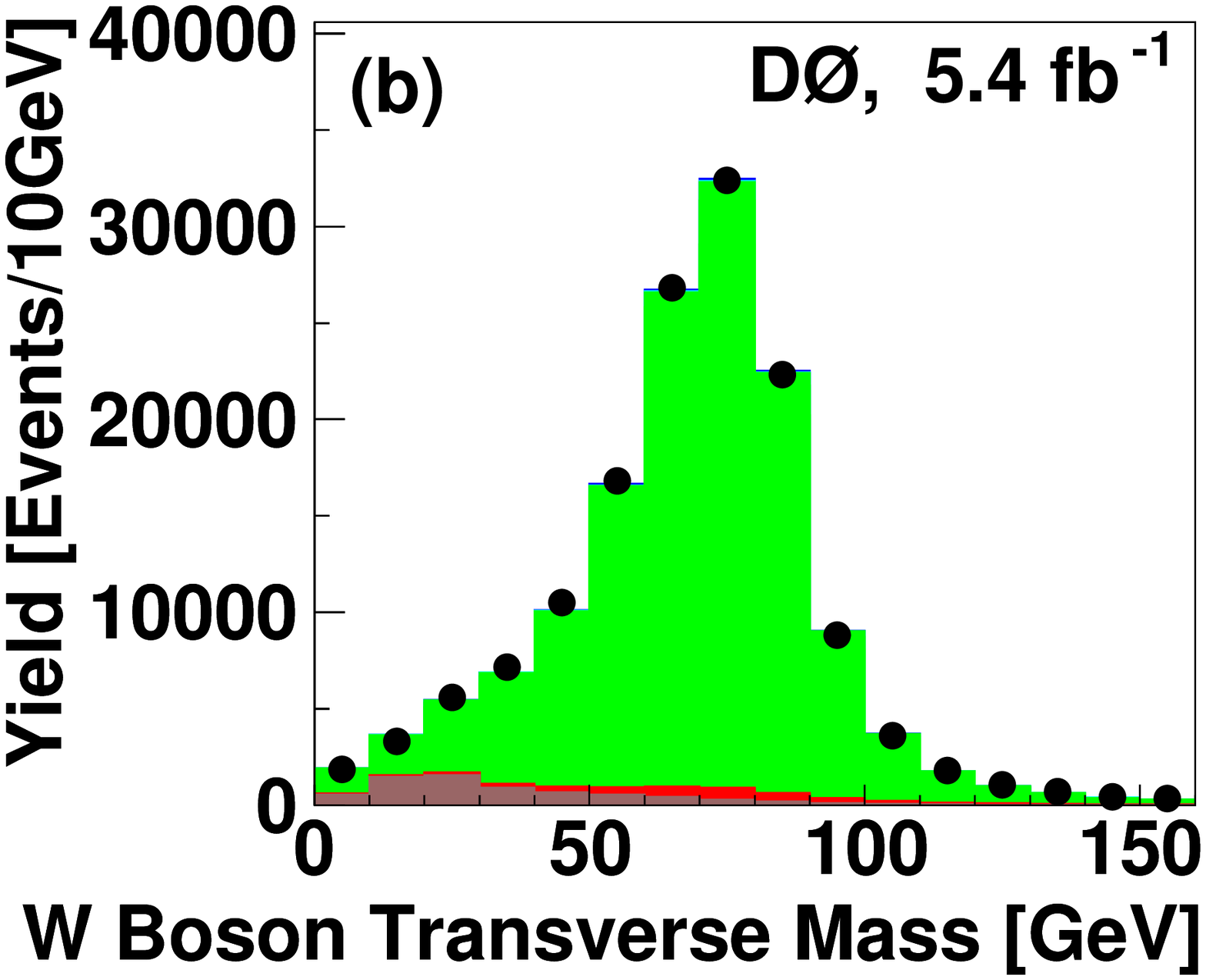}
\includegraphics[width=0.235\textwidth]{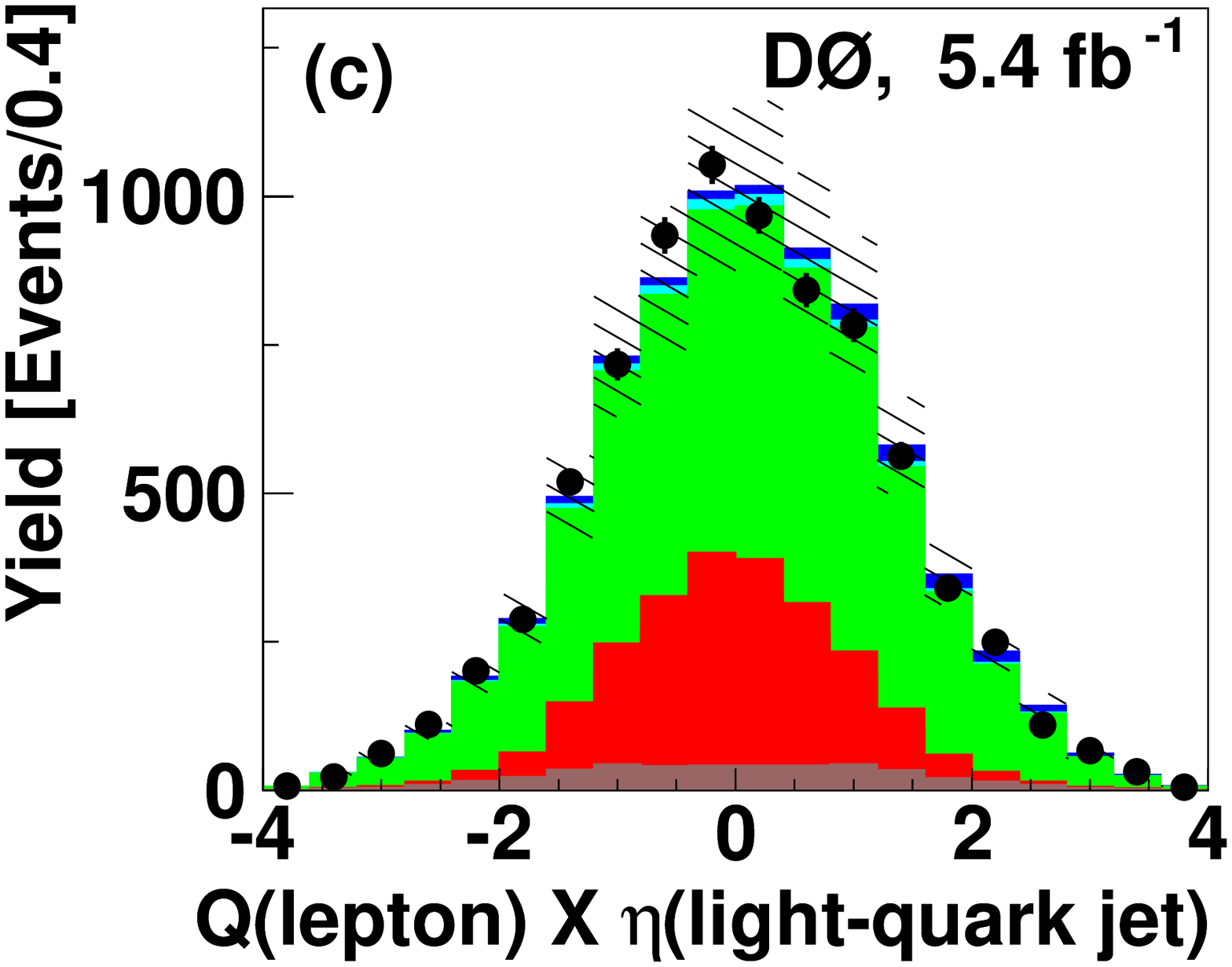}
\includegraphics[width=0.235\textwidth]{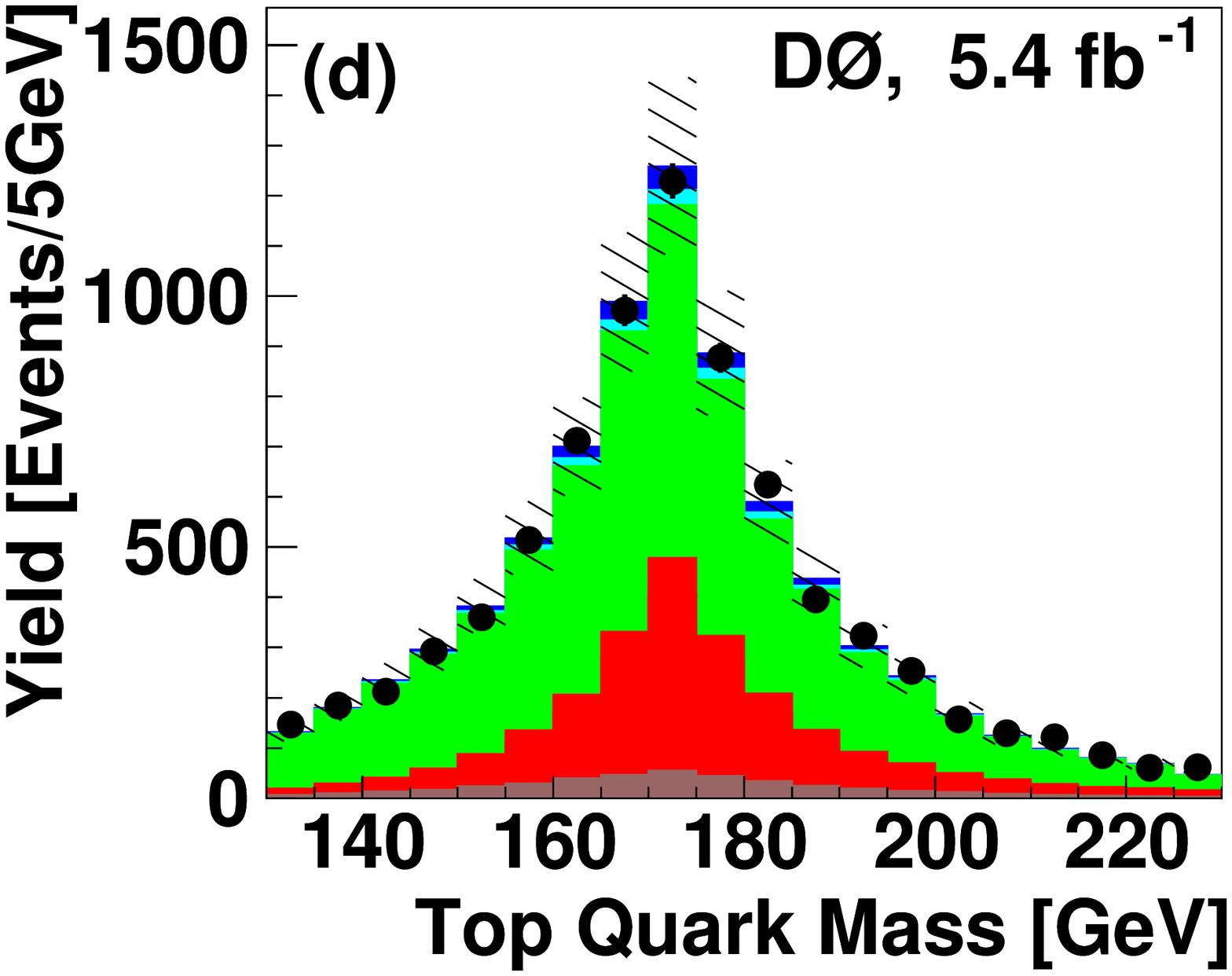}
\includegraphics[width=0.235\textwidth]{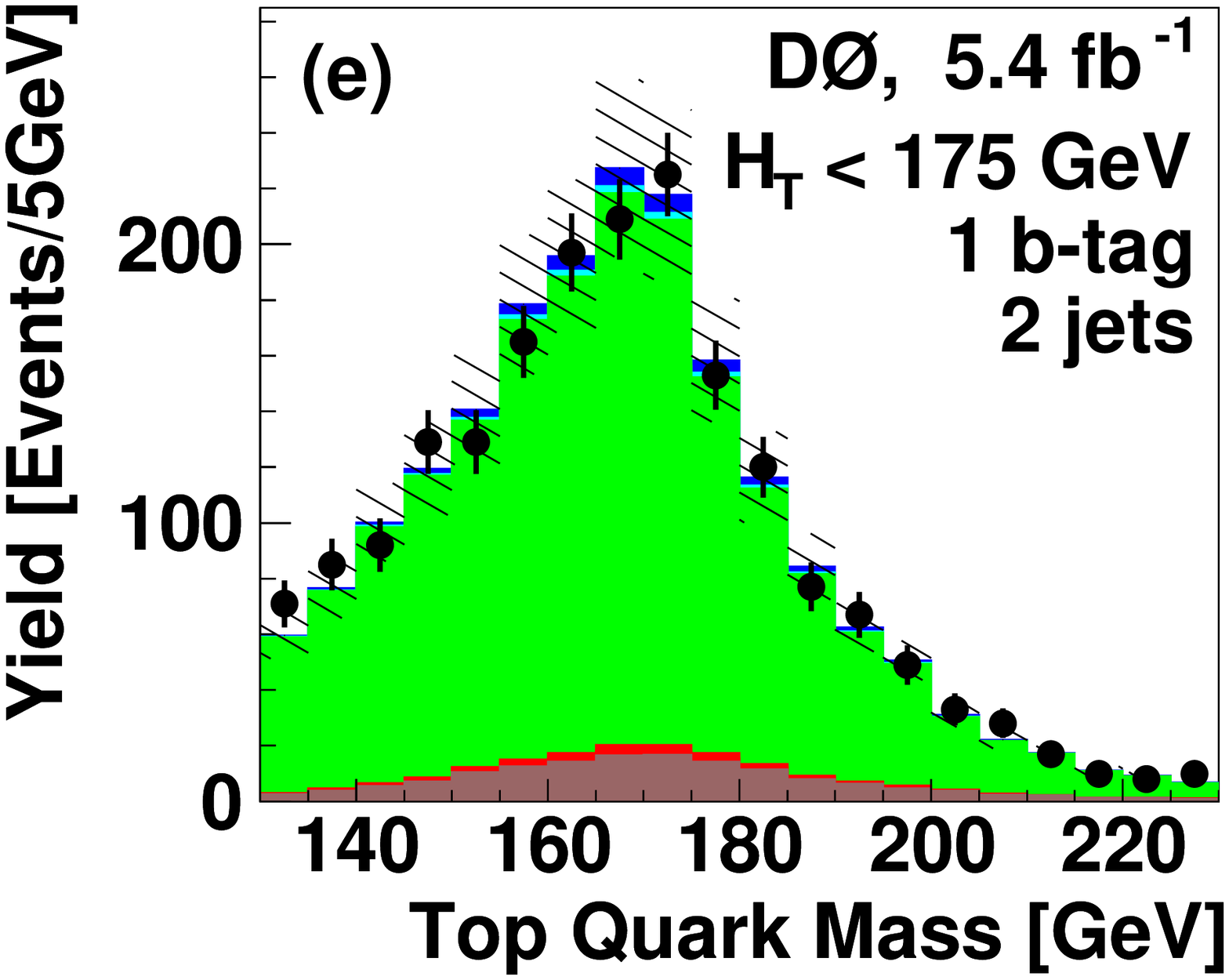}
\includegraphics[width=0.235\textwidth]{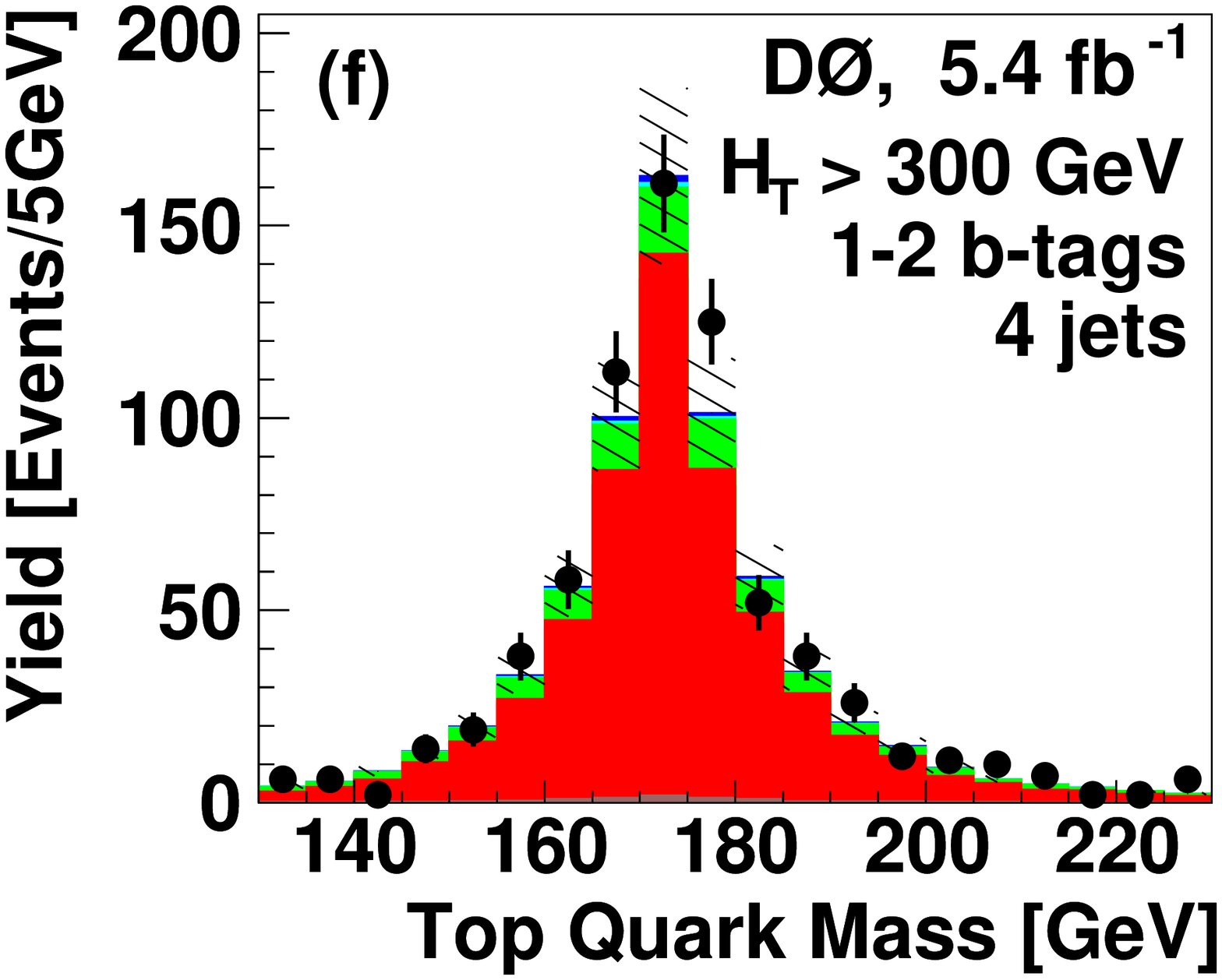}
\caption{[color online]
Comparisons between the data and the background model for (a) $\met$, (b) $W$ boson transverse mass 
before $b$-tagging, and (c) light quark jet pseudorapidity multiplied by lepton charge, after $b$-tagging.  
Reconstructed top quark mass (d) after $b$-tagging, 
(e) in a control sample dominated by $W$+jets, and (f) in a control sample dominated by \ttbar. 
The hatched bands show the $\pm 1\sigma$ uncertainty on the background prediction for  distributions 
obtained after $b$-jet identification (c--f). The $W$+jets contribution includes events 
from $Z$+jets and diboson sources.
} 
\label{fig:variables}
\end{figure}

Table~\ref{tab:event-yields} lists the numbers of expected and observed events for each process after 
event selection, including $b$-tagging. Figure~\ref{fig:variables} shows comparisons between data and 
simulation before and after applying $b$-tagging. In the same figure, the 
normalization and differential spectra of the two dominant backgrounds are checked
using the control samples dominated by $W$+jets (e), and by {\ttbar} (f) events.
These plots are indicative of the adequate background modeling attained for various sample 
conditions in the analysis.

\begin{table}[!h!tbp]
\vspace{-0.15in}
\begin{center}
\caption{Numbers of expected and observed events in a data sample corresponding to $5.4\;\rm fb^{-1}$ of integrated luminosity, 
with uncertainties including both statistical and systematic components. The $tb$ and $tqb$ contributions are 
normalized to their SM expectations for a top quark mass of $172.5\;\rm GeV$.
}
\label{tab:event-yields}
\begin{tabular}{lrclrclrcl}
\hline\hline
Source     & \multicolumn{3}{c}{2 jets} & \multicolumn{3}{c}{3 jets} & \multicolumn{3}{c}{4 jets} \\
\hline
$tb$       & 104 &$\pm$& 16 & 44 &$\pm$& 7.8 & 13 &$\pm$& 3.5 \\
$tqb$      & 140 &$\pm$& 13 & 72 &$\pm$& 9.4 & 26 &$\pm$& 6.4 \\
${\ttbar}$ & 433 &$\pm$& 87 & 830 &$\pm$& 133 & 860 &$\pm$& 163 \\
$W$+jets   & 3,560 &$\pm$& 354 & 1,099 &$\pm$& 169 & 284 &$\pm$& 76 \\
$Z$+jets \& dibosons & 400 &$\pm$& 55 & 142 &$\pm$& 41 & 35 &$\pm$& 18 \\
Multijets  & 277 &$\pm$& 34 & 130 &$\pm$& 17 & 43 &$\pm$& 5.2 \\
\hline
Sum of above sources & 4,914 &$\pm$& 558 & 2,317 &$\pm$& 377 & 1,261 &$\pm$& 272 \\
\hline
Data       & \multicolumn{3}{c}{4,881} & \multicolumn{3}{c}{2,307} & \multicolumn{3}{c}{1,283} \\
\hline\hline
\end{tabular}
\end{center}
\vspace{-0.15in}
\end{table}

\section{Multivariate analyses}

Since the expected single top quark contribution is smaller than the uncertainty on the background, we use multivariate analysis (MVA) 
methods to extract the signal. 
The application of these methods to the measurement of the single top quark production cross section is described
in Ref.~\cite{d0-prd-2008}. 
Three different MVA techniques are used in this analysis: 
(i) Bayesian neural networks (BNN)~\cite{bayesianNNs}, 
(ii) boosted decision trees (BDT)~\cite{decision-trees},  
and (iii) neuroevolution of augmented topologies (NEAT)~\cite{neat}. Each MVA method constructs a function that 
approximates the probability $\text{Pr}(S|\mathbf{x})$ that an event, characterized by the 
variables $\mathbf{x}$, originates from the signal process, $S = \{tb, tqb, tb+tqb\}$. 
Therefore each method defines a discriminant $D$ that can be used to constrain the uncertainties 
of the background in the low-discriminant region $D \approx 0$ and extract a signal from an excess 
in the high-discriminant region $D \approx 1$.
All three methods use the same data and model for background, performing the analyses 
separately on the six mutually exclusive subsamples defined before. All three methods also 
consider the same sources of systematic uncertainty,
and are trained using variables for discriminating signal from background chosen from a common set of
well-modeled variables. These variables can be classified in five categories:
single object kinematics, global event kinematics, jet reconstruction, top quark reconstruction, and angular correlations.
The BNN uses four-vectors of the lepton and jets and a two-vector for $\met$ to build the discriminant. 
The BNN performance is improved by adding variables containing the lepton charge and $b$-tagging information, resulting 
in 14, 18, and 22 variables for events with 2, 3, and 4 jets. 
The BDT ranks and selects the best fifty variables for all the analysis channels, while NEAT uses the 
TMVA~\cite{tmva} implementation of
the ``RuleFit"~\cite{rulefit} algorithm to select the best thirty variables in each channel.

Each MVA method is trained separately for the two single top quark production channels: (i) for the $tb$ discriminants, with $tb$ 
considered signal and $tqb$ treated as a part of the background, and (ii) for $tqb$ discriminants, 
with $tqb$ considered signal and $tb$ treated as a part of the background.

\begin{figure}
\centering
\includegraphics[width=0.238\textwidth]{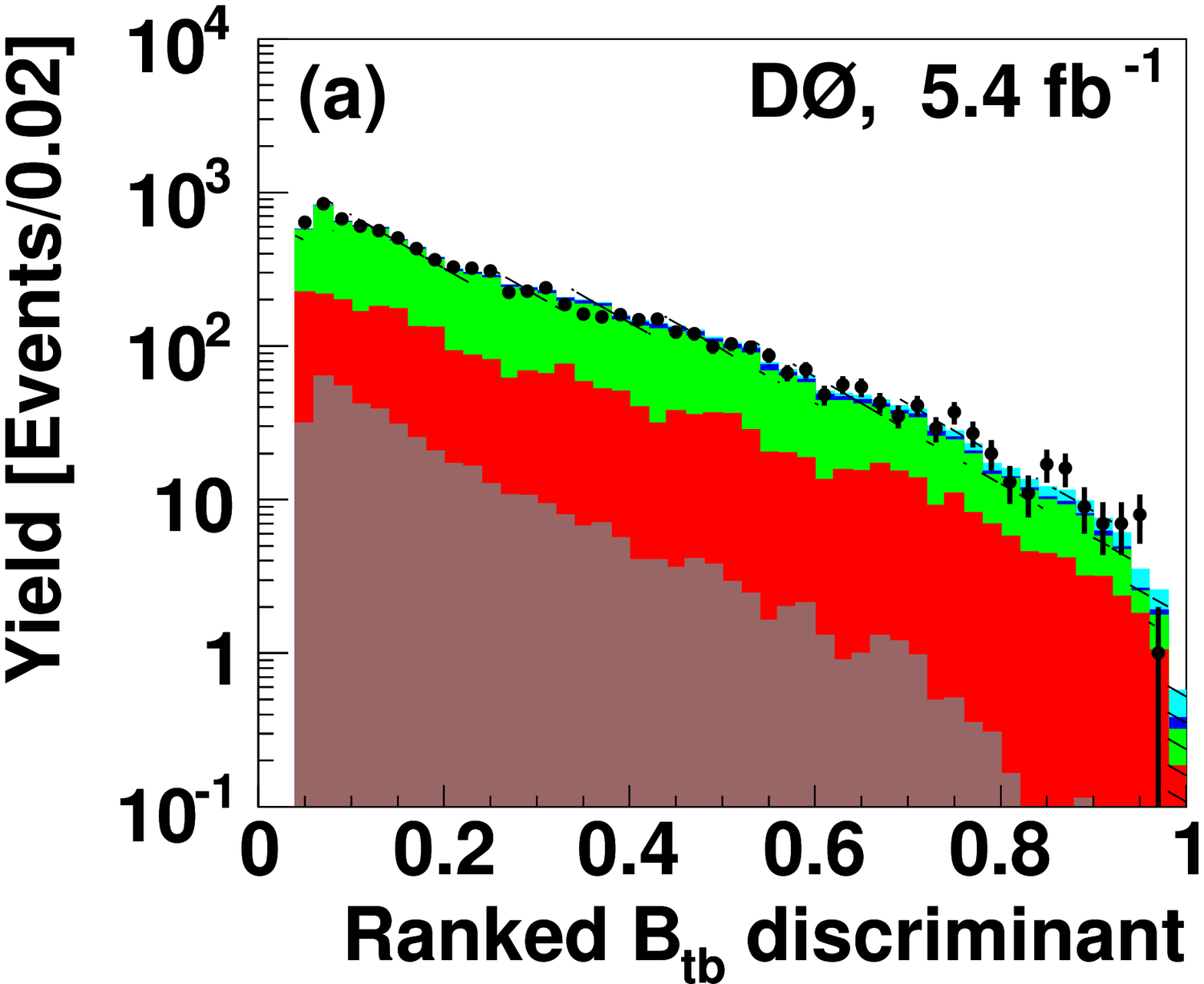}
\includegraphics[width=0.238\textwidth]{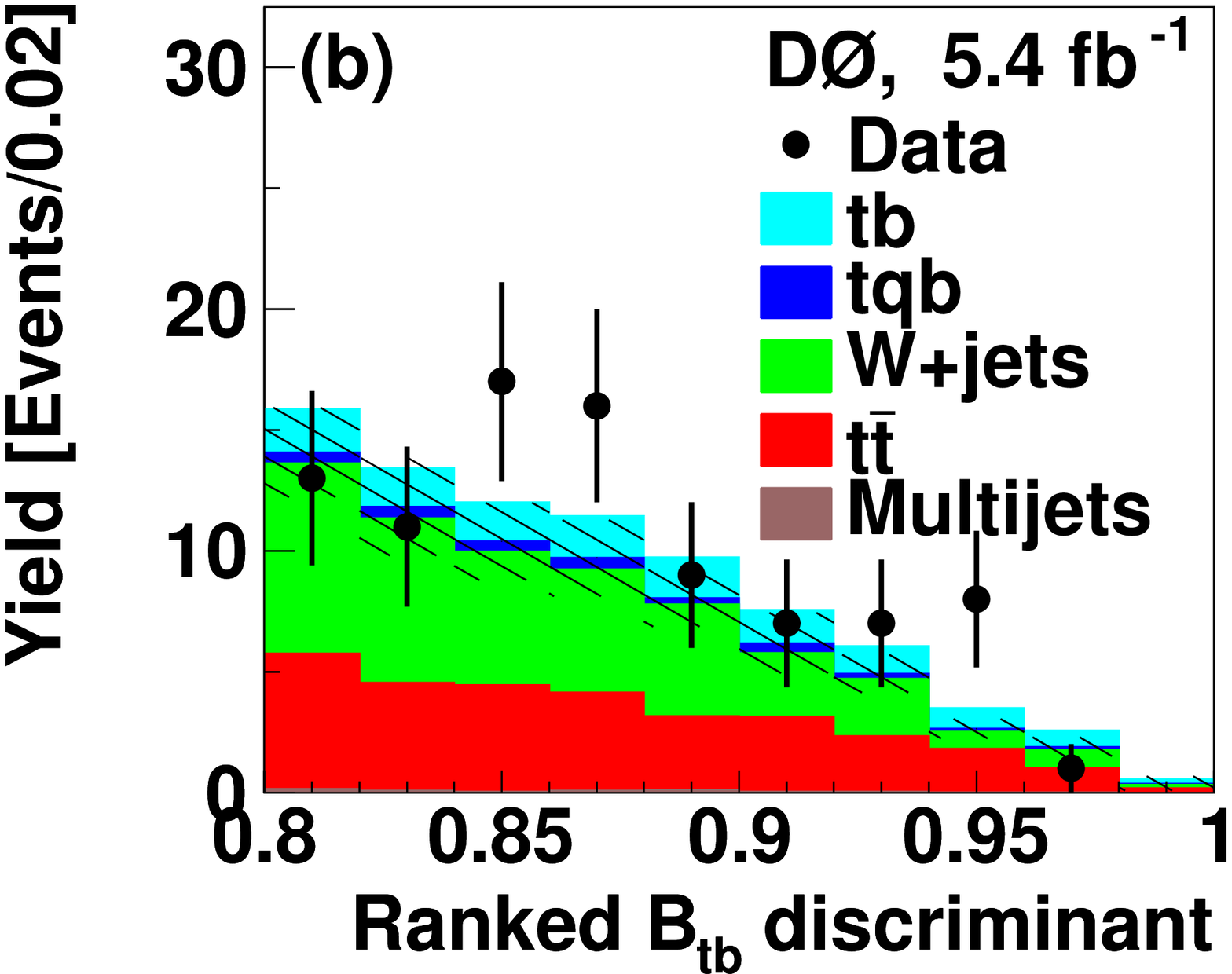}
\includegraphics[width=0.238\textwidth]{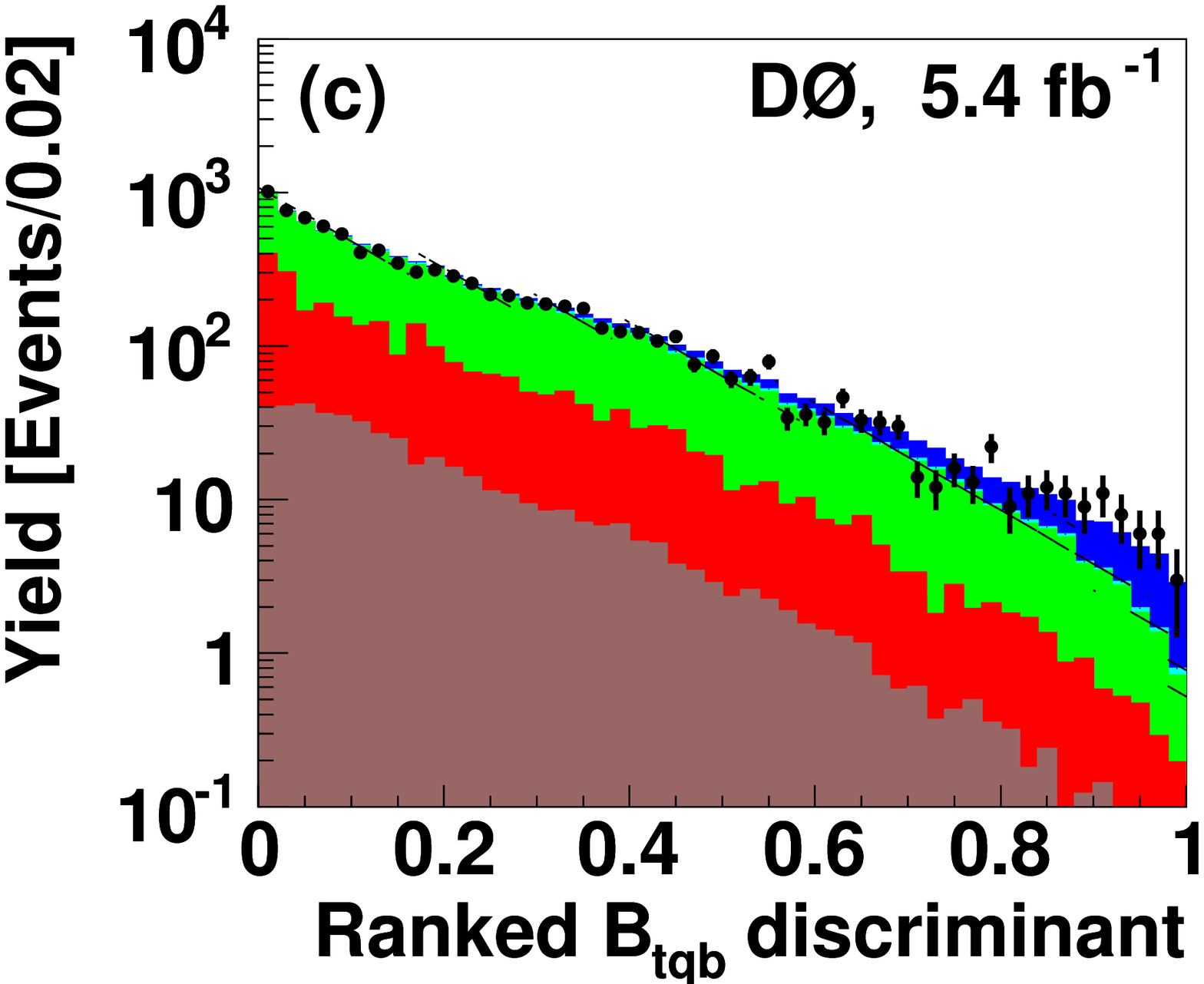}
\includegraphics[width=0.238\textwidth]{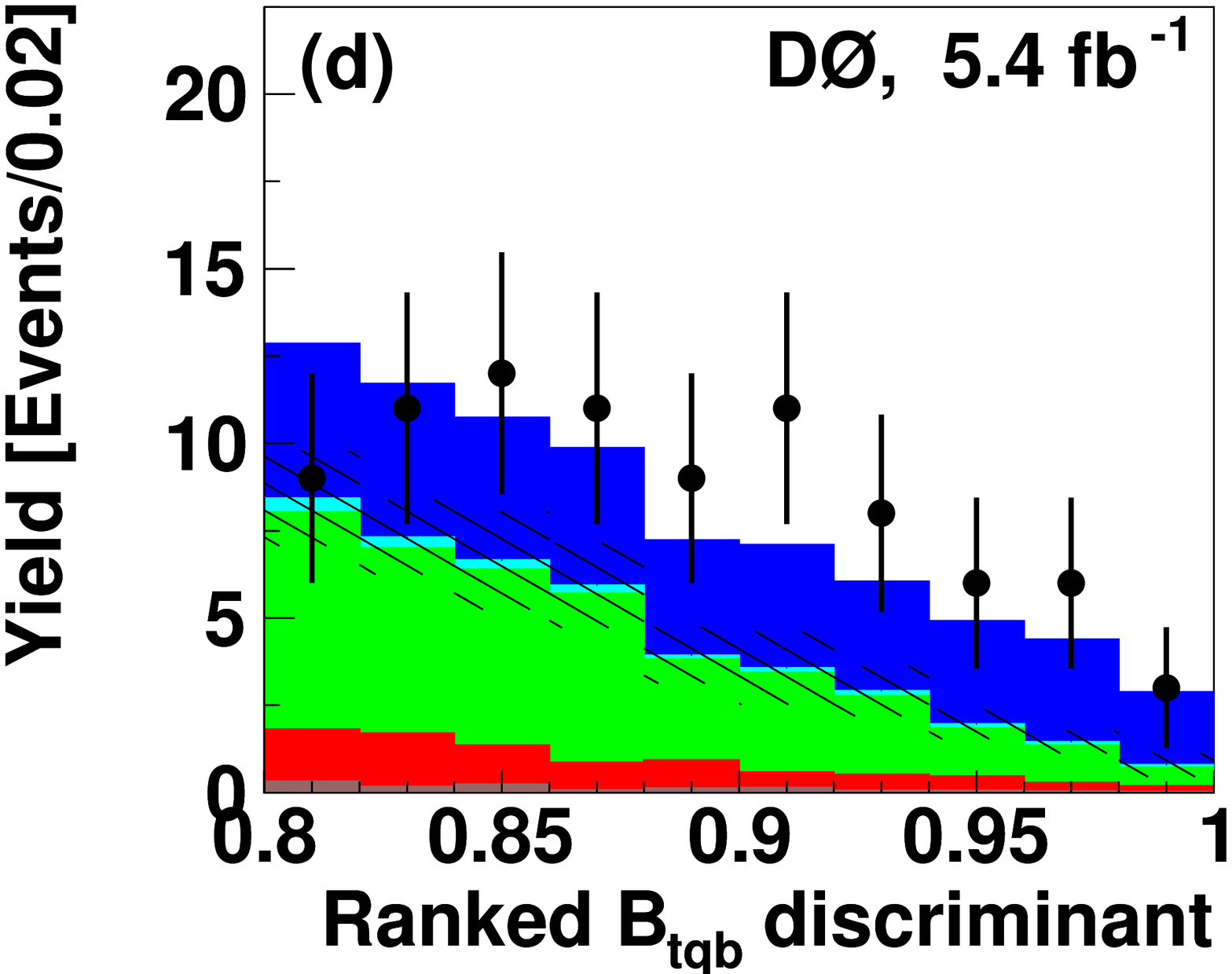}
\includegraphics[width=0.238\textwidth]{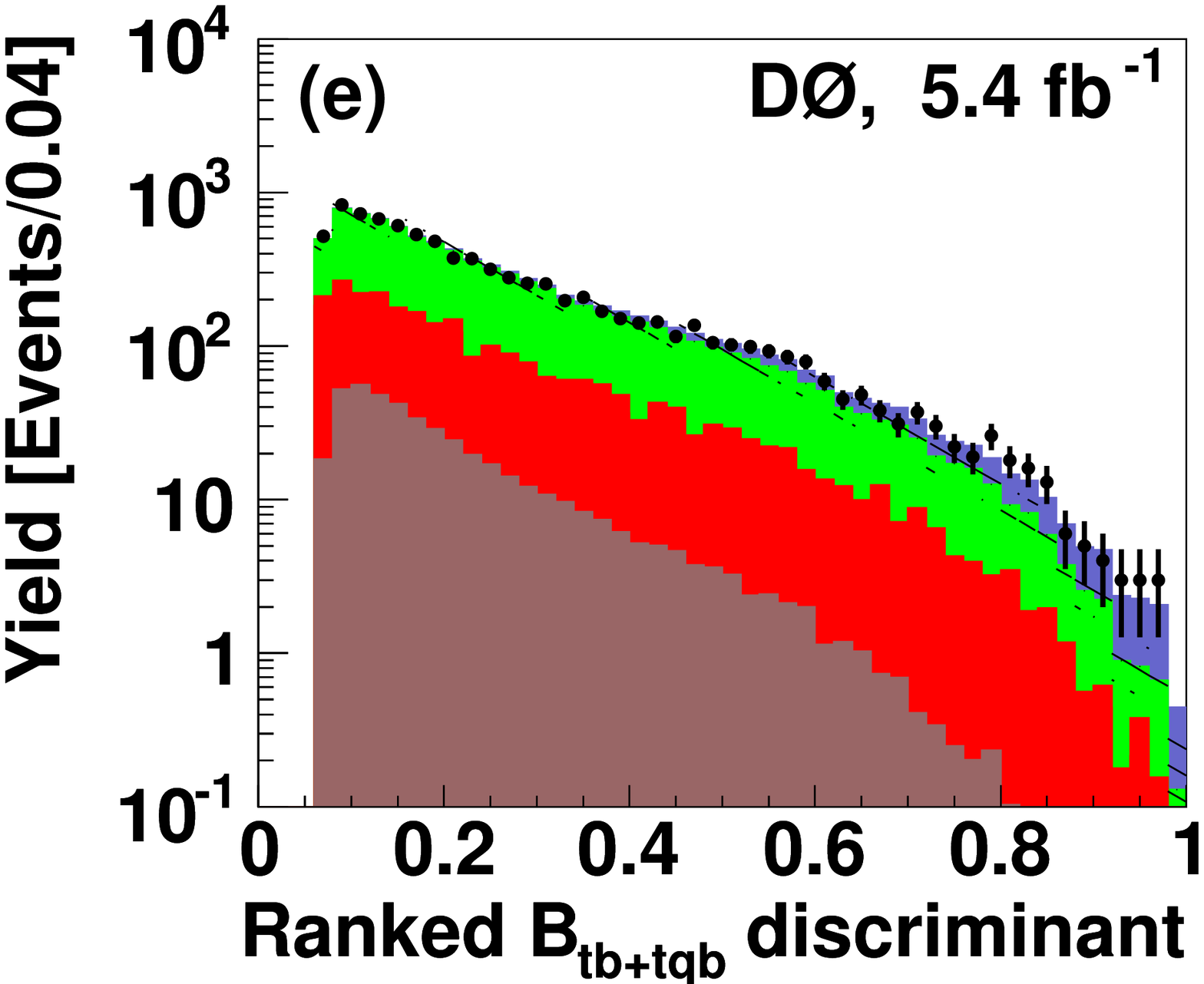}
\includegraphics[width=0.238\textwidth]{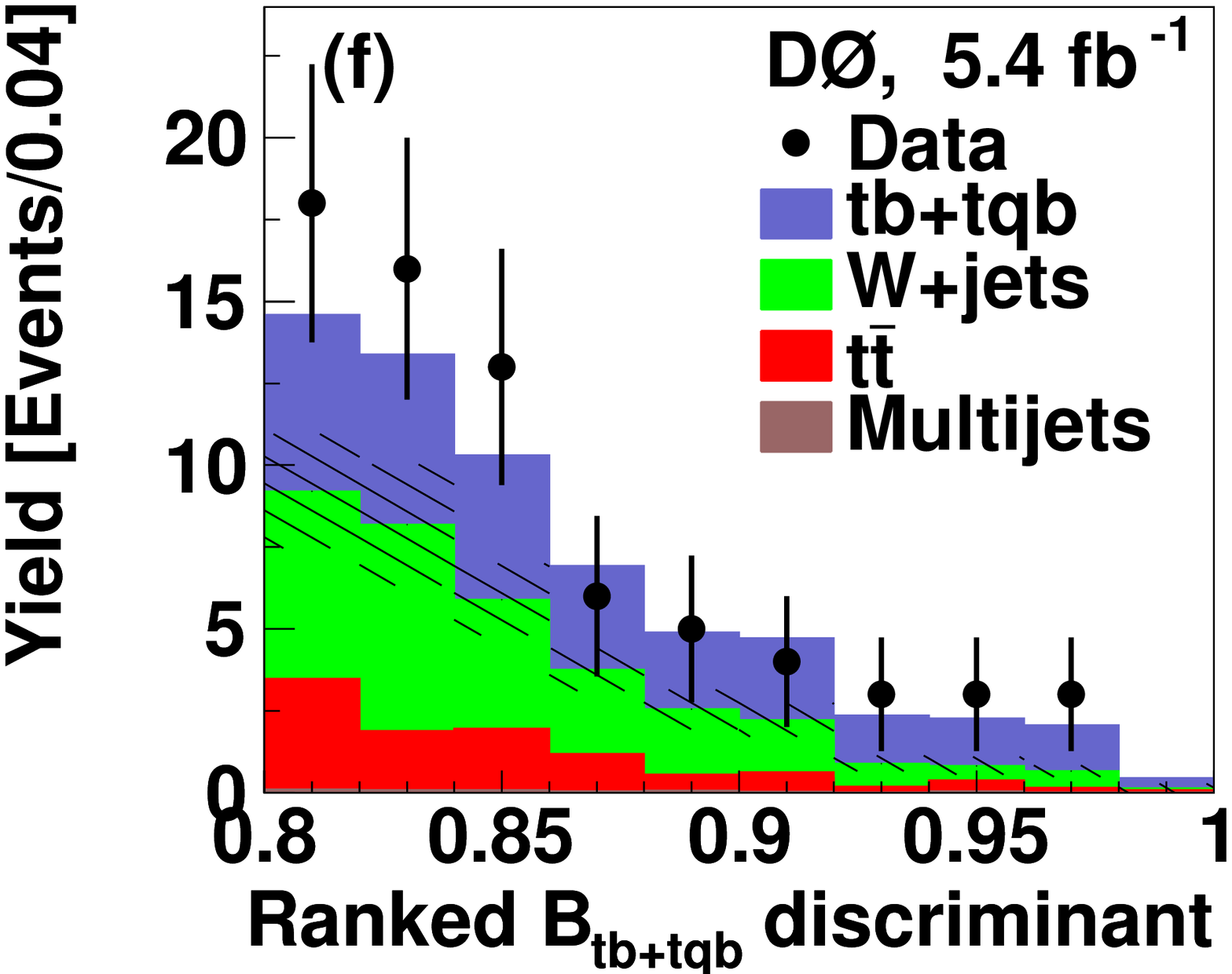}
\vspace{-0.1in}
\caption{[color online] Distributions of the (a) $B_{tb}$, (c) $B_{tqb}$, and (e) $B_{tb+tqb}$ discriminants for the entire range [0--1] of the output.  Distributions of the (b) $B_{tb}$, (d) $B_{tqb}$, and (f) $B_{tb+tqb}$ discriminants for the signal region [0.8--1].
The bins have been ``ranked" by their expected signal-to-background ratio. 
The $tb$, $tqb$, and $tb$+$tqb$ contributions are normalized to the measured cross sections in Table~\ref{tab:xsections}.
The hatched bands show the $\pm 1\sigma$ uncertainty on the background prediction.}
\label{fig:discriminants}
\end{figure}

Using ensembles of datasets containing contributions from background and SM signal, 
we infer that the correlation among the outputs of the 
individual MVA methods is $\approx$~70\%. An increase in sensitivity can therefore be obtained by combining these methods 
to form a new discriminant~\cite{stop-obs-2009-d0}. To achieve the maximum sensitivity,
a second BNN is used to construct a combined discriminant for each channel,
for $tb$, $tqb$, and $tb$+$tqb$ events, defined as $B_{tb}$, $B_{tqb}$ and $B_{tb+tqb}$. 
The $B_{tb}$ and $B_{tqb}$ discriminants take as inputs the three 
discriminant outputs of BDT, BNN, and NEAT, and they are trained by assuming $tb$ or $tqb$ as signals, respectively. 
The combined $tb+tqb$ discriminant ($B_{tb+tqb}$) takes as input the six discriminant 
outputs of BDT, BNN, and NEAT that are trained separately for the $tb$ and the $tqb$ signal. 
The training for $B_{tb+tqb}$ treats the combined $tb$+$tqb$ contribution as signal with relative production 
rates predicted by SM.
Figure~\ref{fig:discriminants} shows the outputs of the $B_{tb}$, $B_{tqb}$, and $B_{tb+tqb}$ discriminants, where good 
agreement is observed over the entire range. In these plots, the bins are sorted and merged (``ranked") 
as a function of the expected signal-to-background ratio (S:B) such that S:B increases monotonically within the range
of the discriminant. For the $tqb$ and $tq$+$tqb$ discriminants, presence of signal is significant in the plots. 
For the $tb$ discriminant, the signal presence is not as significant. 

\section{Measuring signal cross sections}

\subsection{Bayesian approach}

We use a Bayesian approach~\cite{d0-prl-2007, d0-prd-2008, stop-obs-2009-d0} to extract the production cross sections. 
The method consists of forming a binned likelihood as a product of all six analysis channels (2, 3, or 4 
jets with 1 or 2 $b$-tags) and bins using the full discriminant outputs. We assume a Poisson distribution for 
the number of events in each bin and uniform prior probabilities for non-negative values of the signal cross sections 
($tb$, $tqb$ and $tb+tqb$ correspondingly). Systematic uncertainties and their correlations are taken into 
account by integrating over signal acceptances, background yields, and integrated luminosity, assuming a 
Gaussian prior for each source of systematic uncertainty. A posterior probability density as a function of the 
single top quark cross section is constructed, with the position of the maximum defining the value of the cross 
section and the width of the distribution in the region that encompasses 68\% of the entire area corresponding 
to the uncertainty (statistical and systematic components combined). The expected cross sections are obtained 
by setting the number of data events in each channel equal to the value given by the prediction of signal plus 
background.

\subsection{Ensemble testing}

\begin{figure*}
\centering
\includegraphics[width=0.30\textwidth]{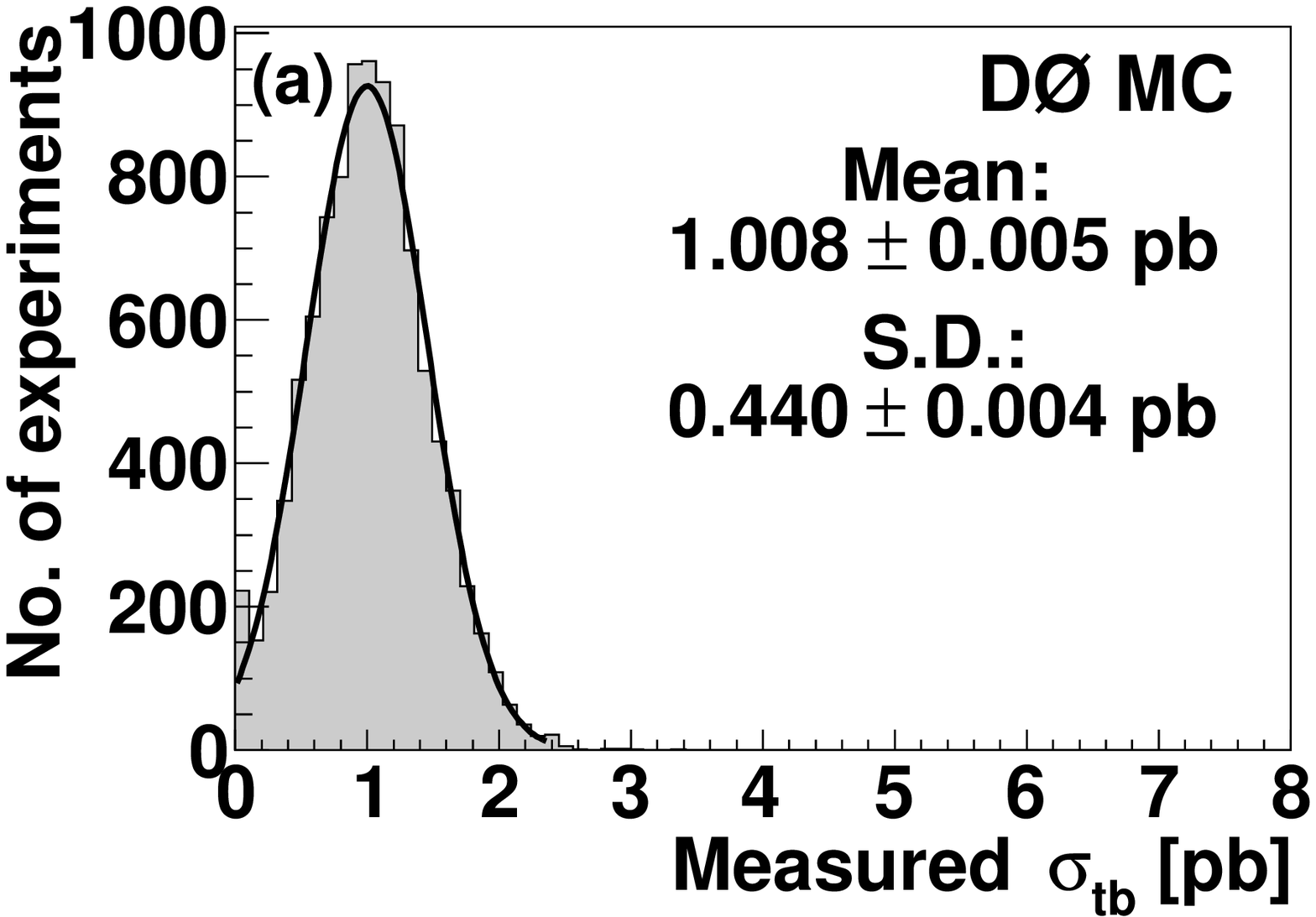}
\includegraphics[width=0.30\textwidth]{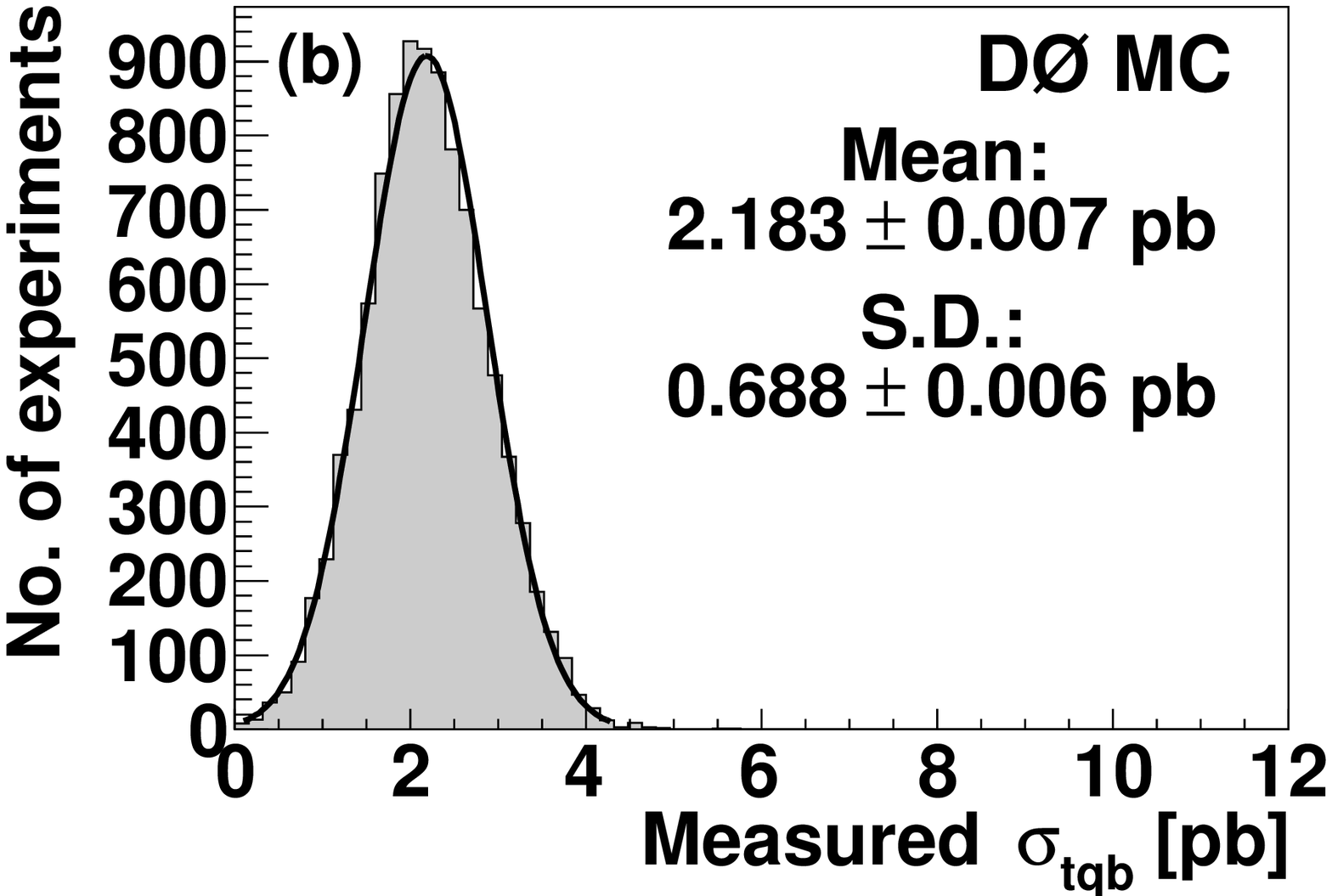}
\includegraphics[width=0.30\textwidth]{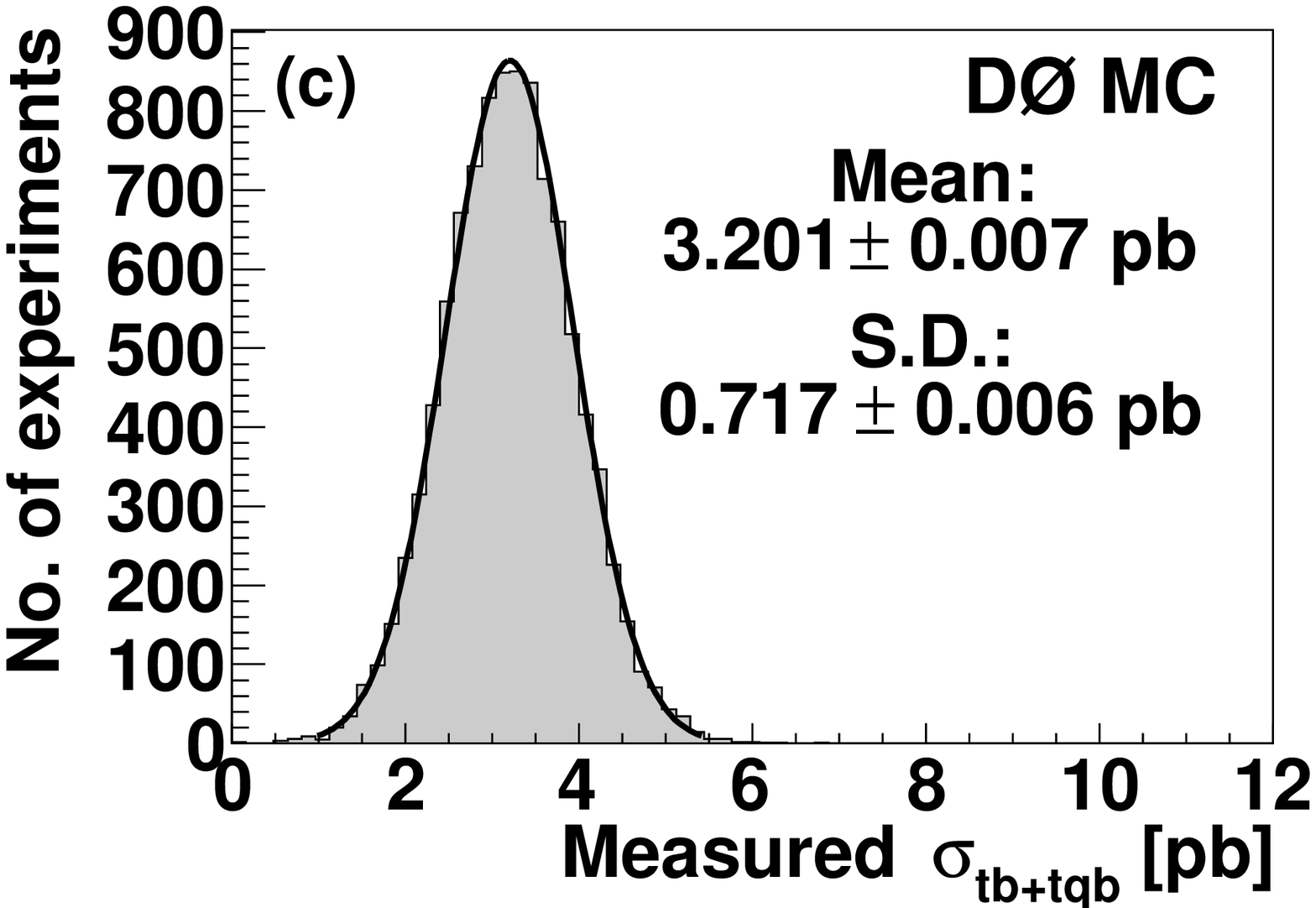}
\vspace{-0.1in}
\caption{Distribution and Gaussian fit of the measured cross section in a ensemble of 
pseudo-experiments with the same integrated luminosity as in data generated assuming the 
SM for (a) $tb$, (b) $tqb$, and (c) $tb+tqb$ processes.
}
\label{fig:gaussian-ensembles}
\end{figure*}

\begin{figure*}
\centering
\includegraphics[width=0.30\textwidth]{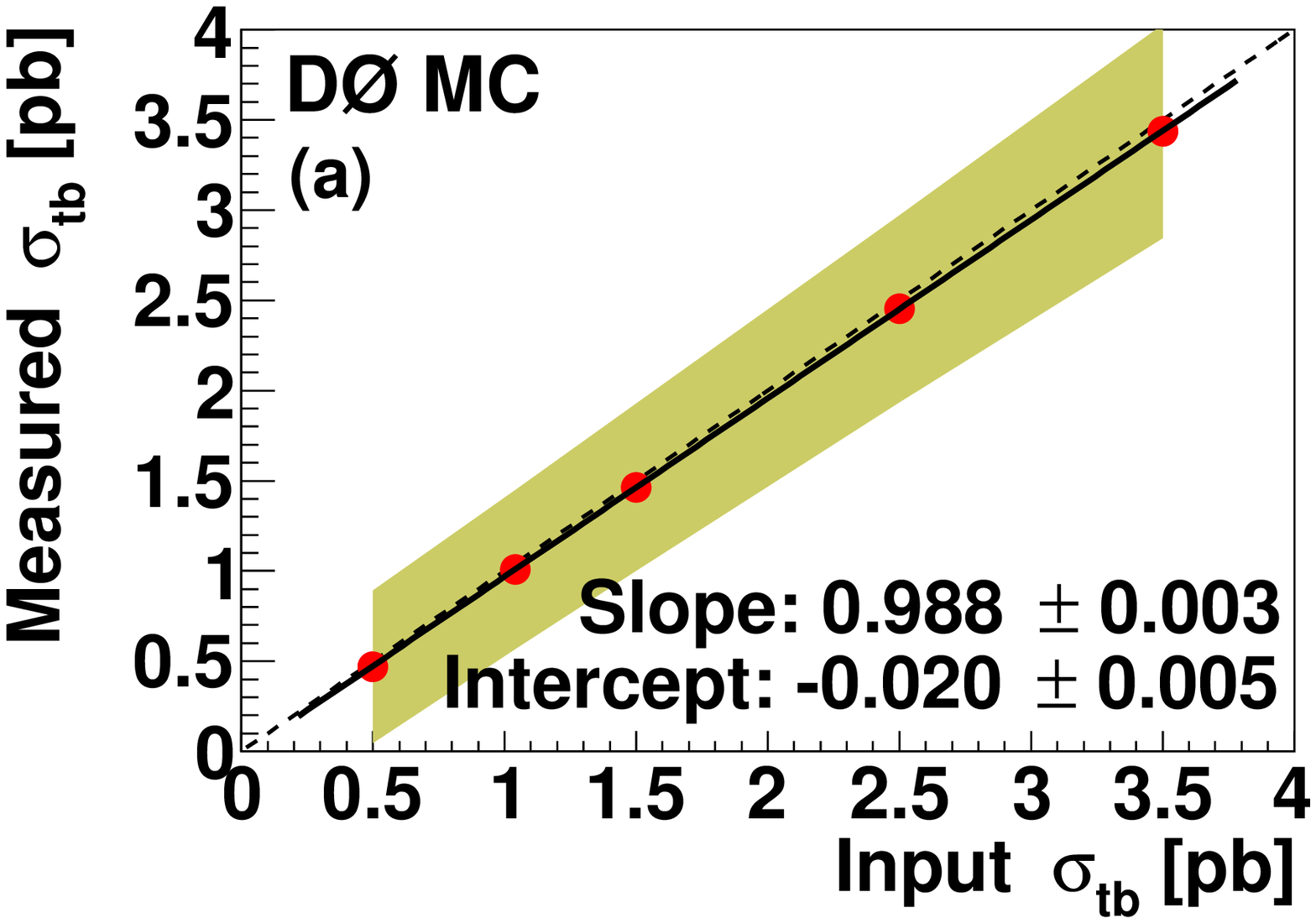}
\includegraphics[width=0.30\textwidth]{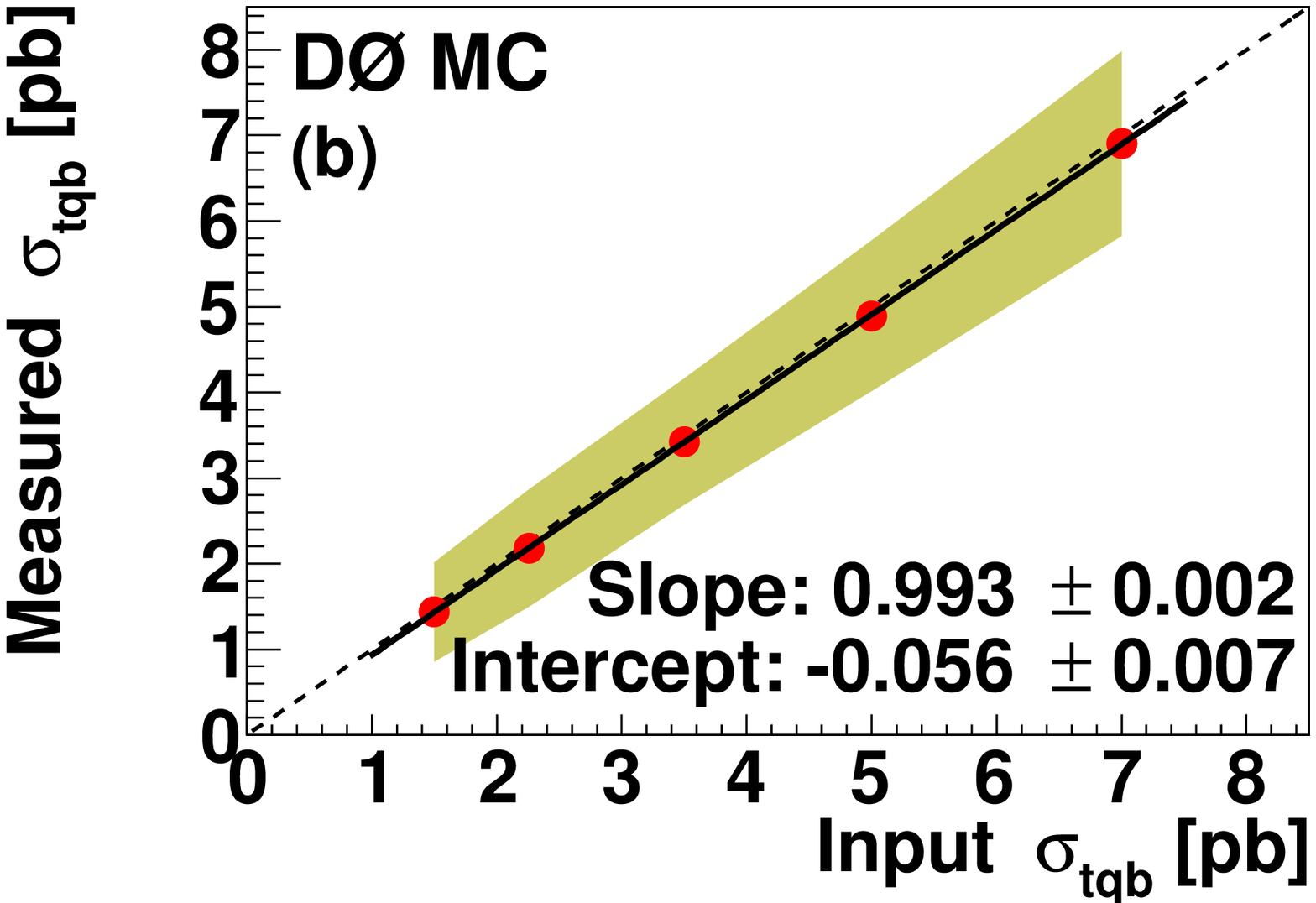}
\includegraphics[width=0.30\textwidth]{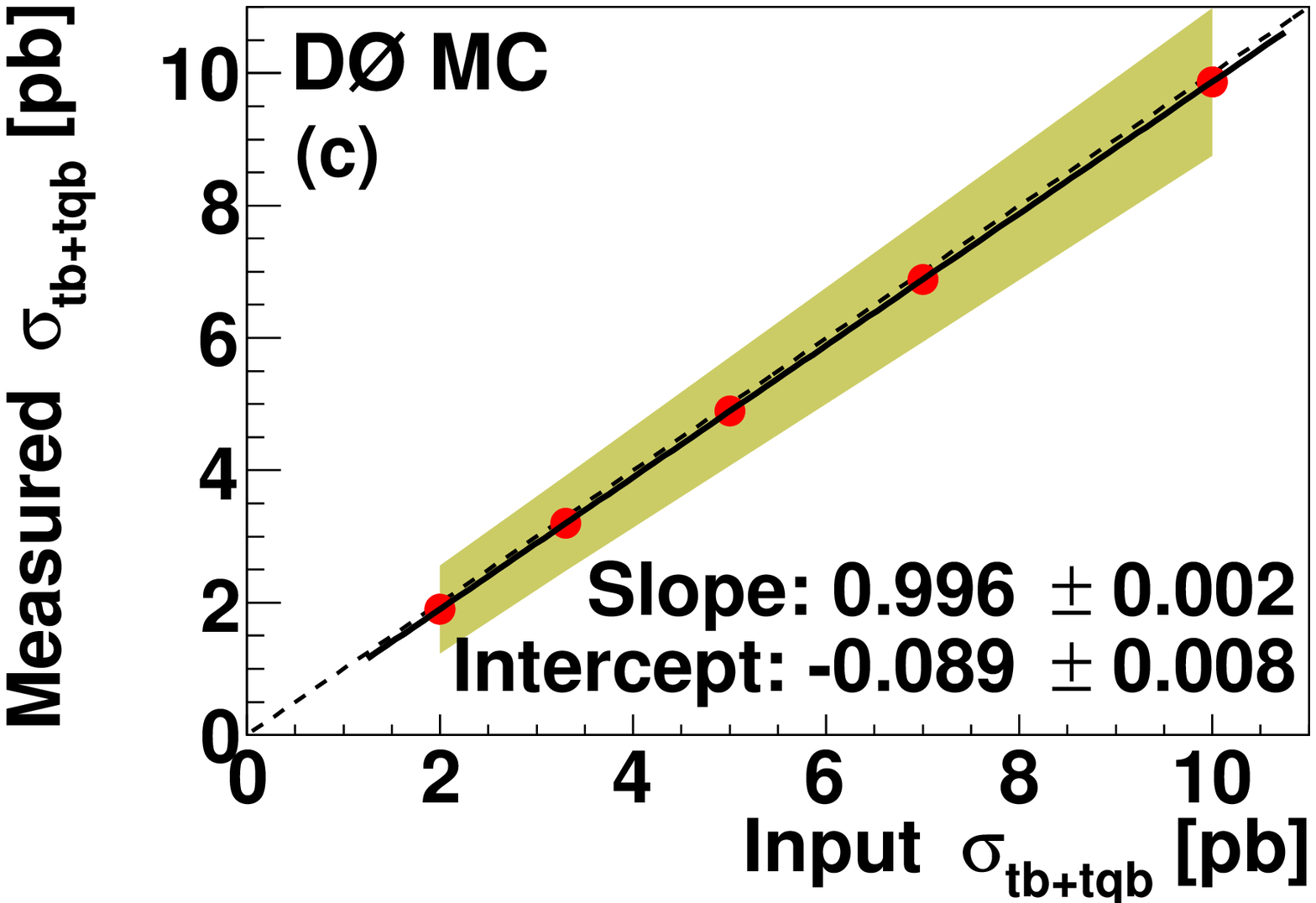}
\vspace{-0.1in}
\caption{[color online] Mean (points) and standard deviation (shaded bands) of cross section as a function of the
input cross section for the (a) $tb$, (b) $tqb$, and (c) $tb$+$tqb$ single top quark processes from the ensemble
studies of pseudo-experiments with the same integrated luminosity as in data.
The continuous lines show the fits to the mean values where their uncertainties
are smaller than the size of the points. The dotted lines represent the responses 
in the case of slope equal one and zero intercept.}
\label{fig:ensembles}
\end{figure*}

The methods used for extracting the cross sections are validated by studies performed using ensembles of pseudo-experiments 
that are generated taking into account all systematic uncertainties and their correlations. 
These ensembles of events are processed
through each MVA method for each single top quark production mode and through the same 
analysis chain as used for the data. Five arbitrary signal cross sections (including the SM 
prediction) are used to calibrate the $tb$, $tqb$, and $tb+tqb$ cross section extraction procedure. 
Means and standard deviations are determined by fitting 
Gaussian function to the distributions of extracted values of the measured cross sections 
in each ensemble. Figure~\ref{fig:gaussian-ensembles} shows the resulting distributions and Gaussian 
fits for SM ensembles for $tb$, $tqb$, and $tb+tqb$ processes.
Straight-line fits
of the extracted mean cross sections to the input values are shown in Fig.~\ref{fig:ensembles}, where 
the shaded bands reflect the standard deviations of the extracted cross sections in each ensemble.

The results of these pseudo-experiments show that the biases on the cross sections are negligible compared to 
the standard deviations of the extracted values. We therefore do not apply corrections to the measured values of the cross sections in data.

\subsection{$\boldsymbol{tb}$, $\boldsymbol{tqb}$ and $\boldsymbol{tb+tqb}$ channel cross sections}

To measure the individual $tb$ ($tqb$) production cross section, we construct a one-dimensional (1D) posterior 
probability density function with the $tqb$ ($tb$) contribution normalized with a Gaussian prior centered on the 
predicted SM cross section and treated as a part of the background. 
This is implemented for each individual MVA method and also for their combination. 
To measure the total single top quark production cross section of $tb$+$tqb$, 
we construct a 1D posterior probability density function assuming the production 
ratio of $tb$ and $tqb$ predicted by the SM.

\begin{figure*}[!ht]
\centering
\includegraphics[width=0.30\textwidth]{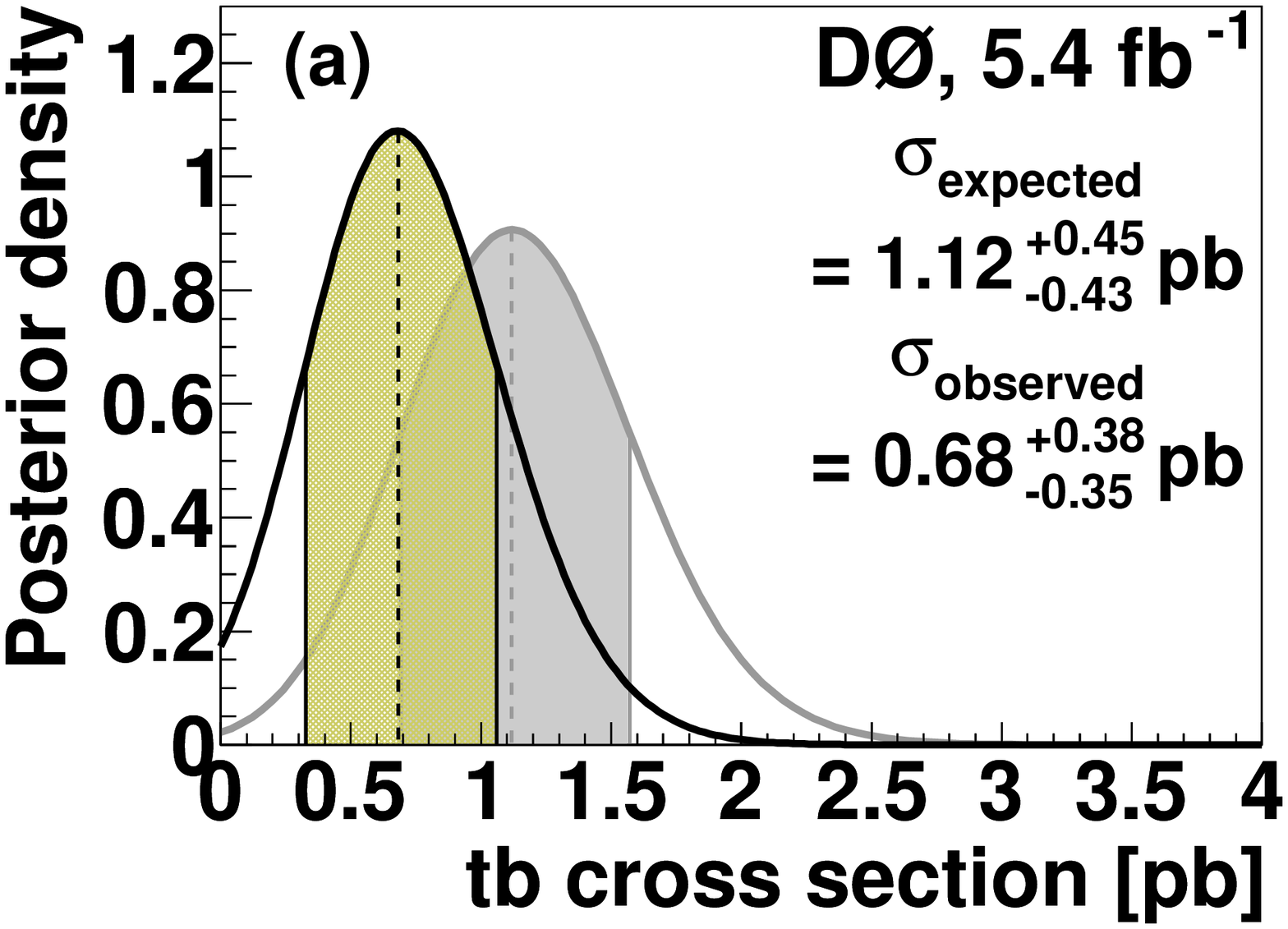}
\includegraphics[width=0.30\textwidth]{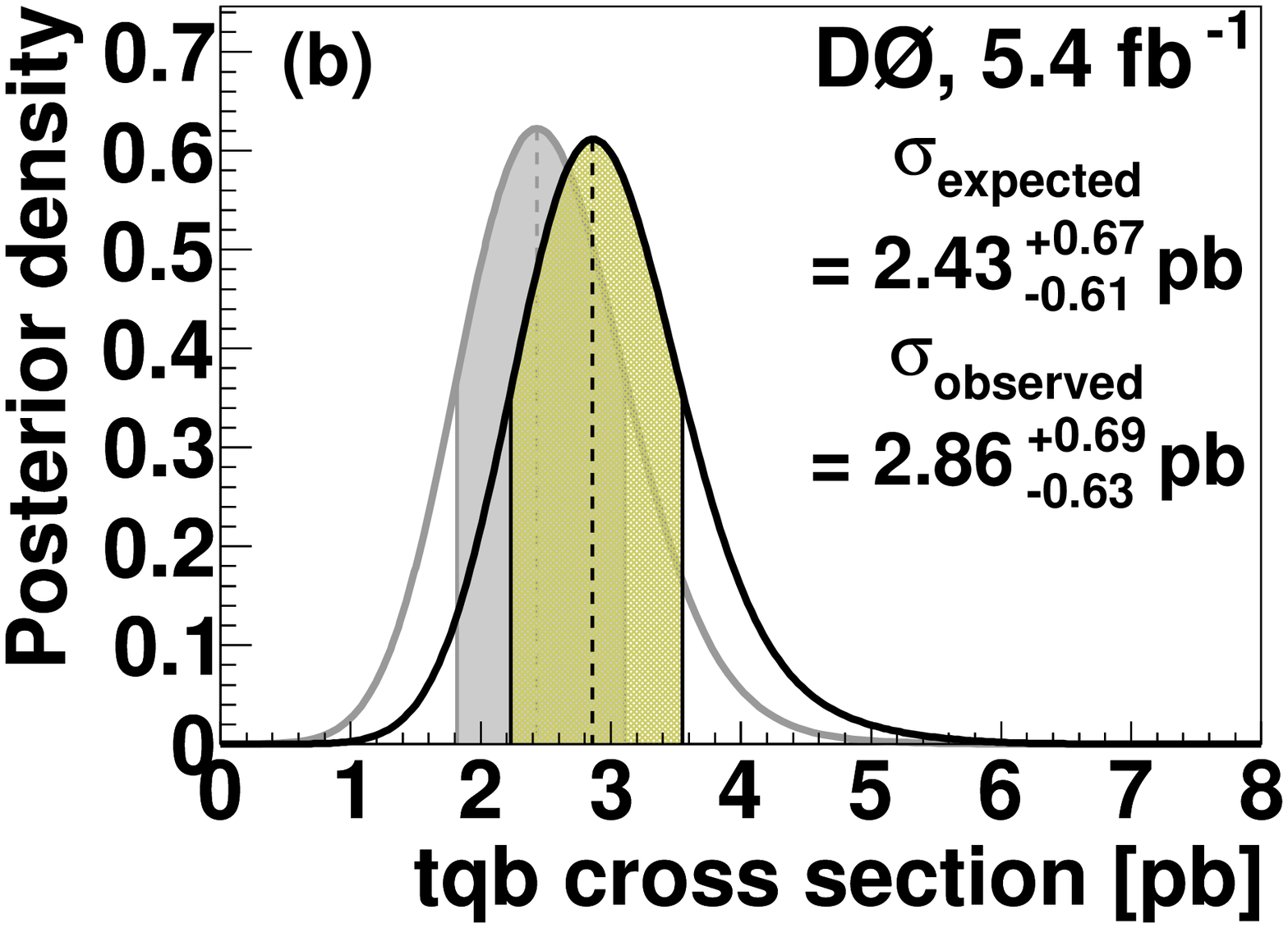}
\includegraphics[width=0.30\textwidth]{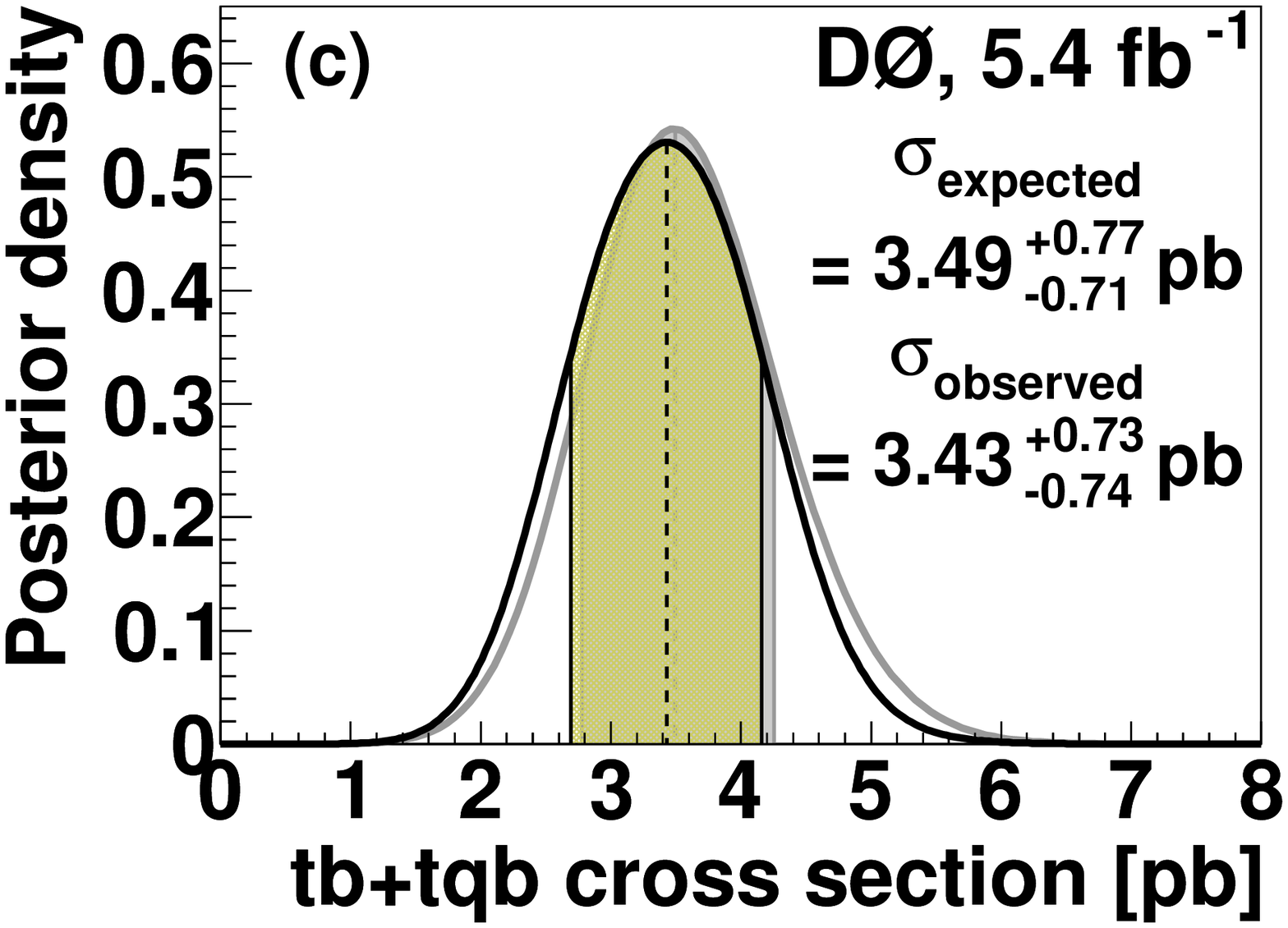}
\vspace{-0.1in}
\caption{[color online] The expected (back) and observed (front) posterior probability densities 
for (a) $tb$, (b) $tqb$, and (c) $tb$+$tqb$ production. 
The shaded bands indicate the 68\% C.L.s from the peak values.} 
\label{fig:posterior1d}
\end{figure*}

Figure~\ref{fig:posterior1d} shows the expected and observed posterior density distributions 
for $tb$, $tqb$, and $tb+tqb$ using the combined discriminants $B_{tb}$, $B_{tqb}$, and $B_{tb+tqb}$, 
respectively. 
Table~\ref{tab:xsections} lists the expected and measured cross sections for the individual MVA analyses. 
All of the results are consistent with SM predictions, and the measured $tb$+$tqb$ production cross section 
is the most precise current measurement, with a precision comparable to the 
world average~\cite{TeVComb}. All results assume a top quark mass of $172.5\;\rm GeV$
and have a small correction for events with more than four jets based on the SM.
The dependence of the measured cross section on $m_t$ is summarized in Table~\ref{tab:topmass}.
The assumed top quark mass affects the yield and differential properties
for the signal acceptance and the modeling of {\ttbar} events, which constitute the second largest background.
The interplay between these two effects can cause the measured cross section to vary substantially (as observed in the $tb$ channel) or in a way that is not monotonic with the assumed top quark mass (as observed in the $tqb$ channel).

\begin{table}[!h!tbp]
\begin{center}
\caption{Expected and observed cross sections in pb for $tb$, $tqb$, and $tb$+$tqb$ production. 
All results assume a top quark mass of $172.5\;\rm GeV$.}
\label{tab:xsections}
\begin{ruledtabular}
\begin{tabular}{lcc}
Discriminant & Expected & Observed \\
\hline
\multicolumn{3}{c}{\underline{$tb$ production}} \\
BNN & $1.08^{+0.52}_{-0.50}$ & $0.72^{+0.44}_{-0.43}$ \\ 
BDT & $1.07^{+0.47}_{-0.43}$ & $0.68^{+0.41}_{-0.39}$ \\
NEAT & $1.06^{+0.54}_{-0.50}$ & $0.17^{+0.41}_{-0.17}$ \\
$B_{tb}$ & $1.12^{+0.45}_{-0.43}$ & $0.68^{+0.38}_{-0.35}$ \\
\multicolumn{3}{c}{\underline{$tqb$ production}} \\
BNN & $2.49^{+0.76}_{-0.67}$ & $2.92^{+0.87}_{-0.73}$ \\ 
BDT & $2.40^{+0.71}_{-0.66}$ & $3.03^{+0.78}_{-0.66}$  \\
NEAT & $2.36^{+0.80}_{-0.77}$ & $2.75^{+0.87}_{-0.75}$ \\
$B_{tqb}$ & $2.43^{+0.67}_{-0.61}$ & $2.86^{+0.69}_{-0.63}$ \\
\multicolumn{3}{c}{\underline{$tb+tqb$ production}} \\
BNN & $3.46^{+0.84}_{-0.78}$ & $3.11^{+0.77}_{-0.71}$ \\ 
BDT & $3.41^{+0.82}_{-0.74}$ & $3.01^{+0.80}_{-0.75}$  \\
NEAT & $3.33^{+0.94}_{-0.80}$ & $3.59^{+0.96}_{-0.80}$ \\
$B_{tb+tqb}$  & $3.49^{+0.77}_{-0.71}$ & $3.43^{+0.73}_{-0.74}$ 
\end{tabular}
\end{ruledtabular}
\end{center}
\end{table}

\begin{table}[!ht]
\begin{center}
\caption{Dependence on $m_t$ of the measured cross sections in pb for $tb$, $tqb$, and $tb$+$tqb$ production, 
using the combined discriminants for the assumed top quark masses. 
The predicted cross sections~\cite{singletop-xsec-kidonakis} in pb are also included in the table and labeled ``SM''.}
\label{tab:topmass}
\begin{ruledtabular}
\begin{tabular}{lccc}
$m_t$   & 170 GeV  & 172.5 GeV & 175 GeV \\
\hline
$tb$ & $1.20^{+0.62}_{-0.56}$ & $0.68^{+0.38}_{-0.35}$ & $0.53^{+0.36}_{-0.34}$ \\
SM & $1.12^{+0.04}_{-0.04}$ & $1.04^{+0.04}_{-0.04}$ & $0.98^{+0.04}_{-0.04}$ \\ \hline
$tqb$ & $2.65^{+0.65}_{-0.59}$ & $2.86^{+0.69}_{-0.63}$ & $2.45^{+0.60}_{-0.57}$ \\
SM & $2.34^{+0.12}_{-0.12}$ & $2.26^{+0.12}_{-0.12}$ & $2.16^{+0.12}_{-0.12}$ \\ \hline
$tb$+$tqb$ & $3.70^{+0.78}_{-0.80}$ & $3.43^{+0.73}_{-0.74}$ & $2.56^{+0.69}_{-0.61}$ \\
SM & $3.46^{+0.16}_{-0.16}$ & $3.30^{+0.16}_{-0.16}$ & $3.14^{+0.16}_{-0.16}$ 
\end{tabular}
\end{ruledtabular}
\end{center}
\end{table}

\section{Signal dominated distributions}

\begin{figure*}[!htpb]
\centering
\includegraphics[width=0.235\textwidth]{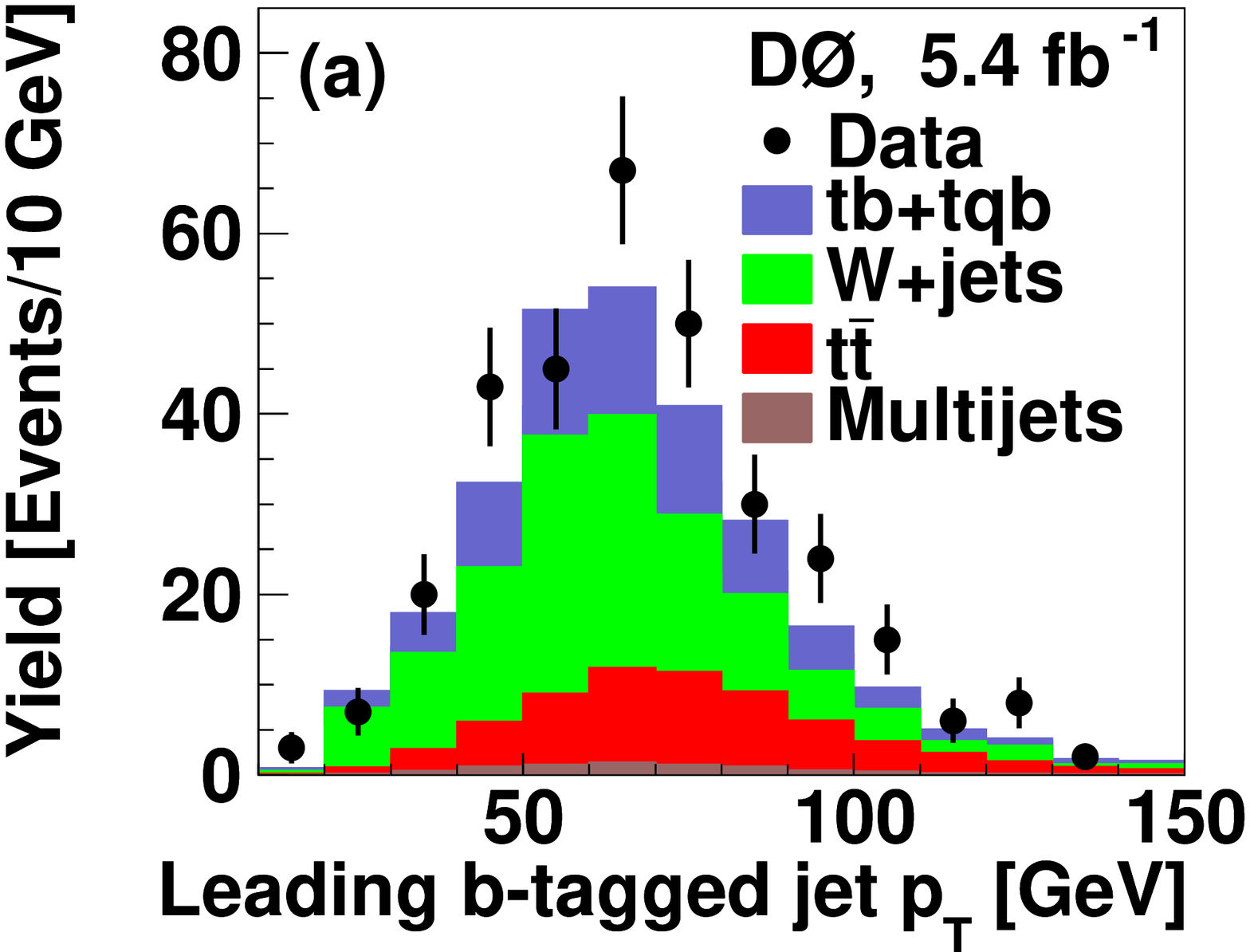}
\includegraphics[width=0.235\textwidth]{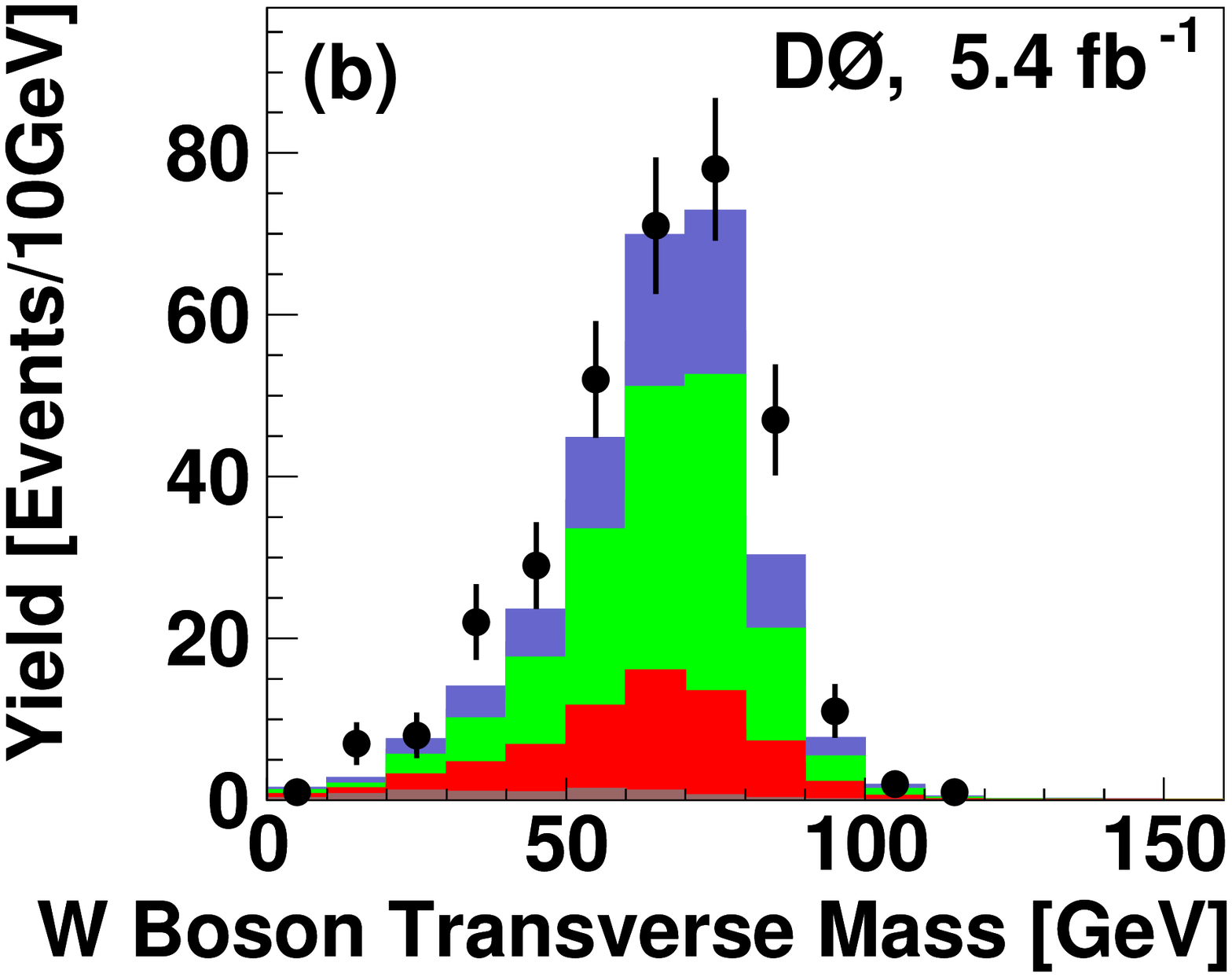}
\includegraphics[width=0.235\textwidth]{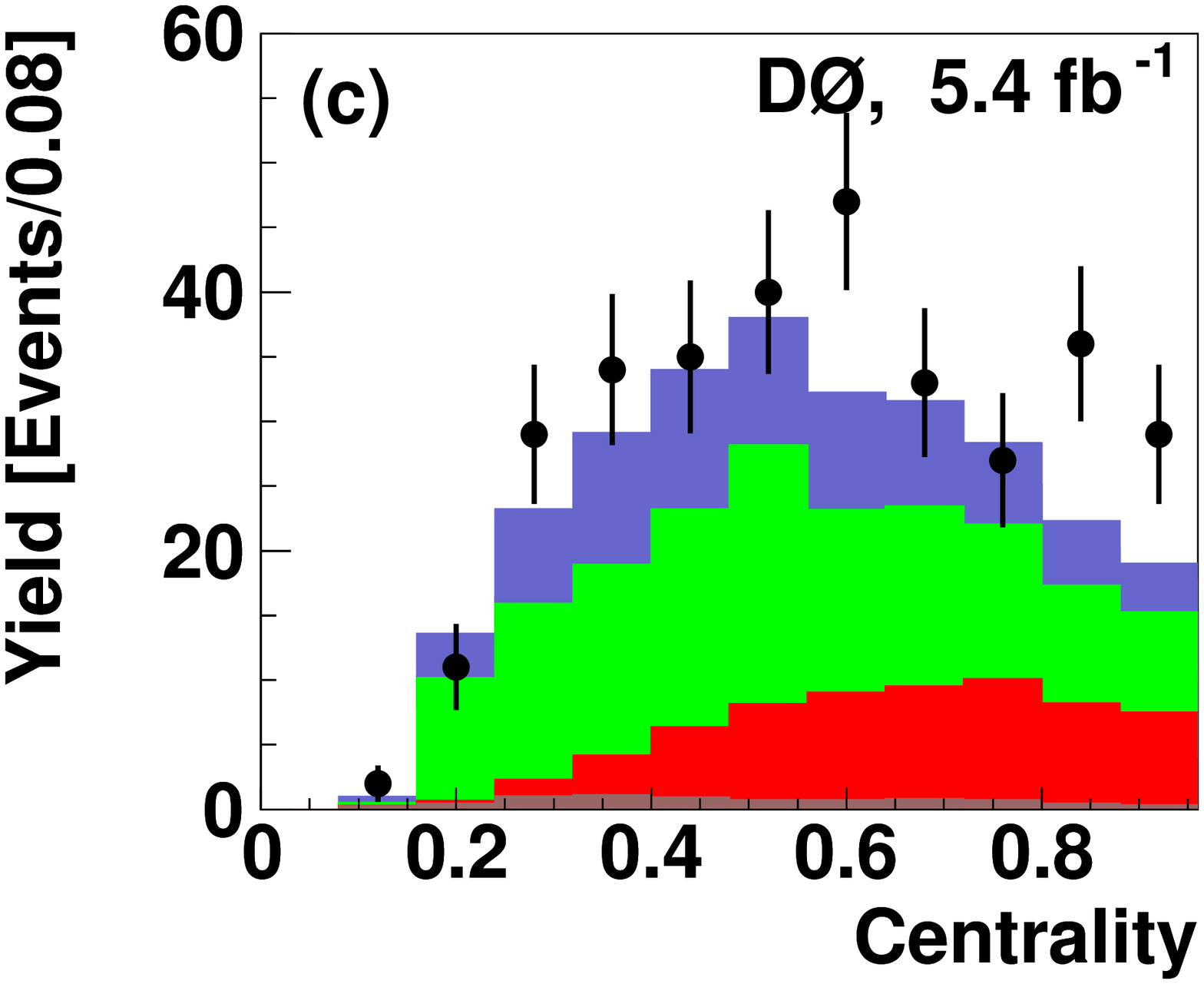}
\includegraphics[width=0.235\textwidth]{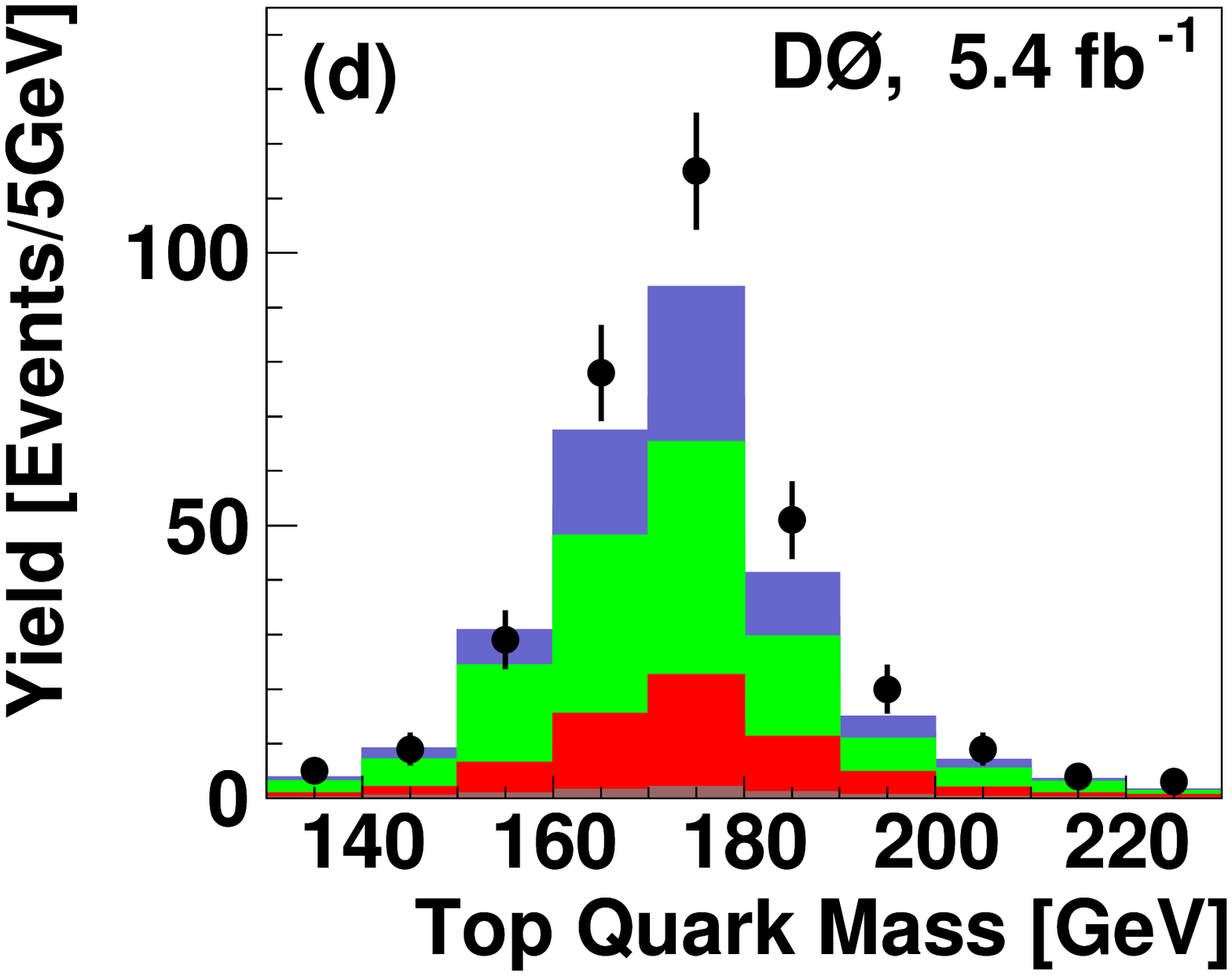}
\caption{[color online] Distributions for data in the regions of large value for signal discrimination: (a) leading $b$-tagged jet $p_T$, (b) $W$ boson transverse mass, (c) centrality, defined as the ratio of the scalar sum of the $p_T$ of the jets
to the scalar sum of the energy of the jets in the event, and (d) reconstructed $m_t$. 
The contributions from signal have been normalized to the measured $tb$+$tqb$ cross section.}
\label{fig:distributions}
\end{figure*}

Figure~\ref{fig:distributions} shows a comparison of the distributions of four kinematic variables with large discriminating power,
for single top quark production in a data sample selected with S:B~$>0.24$ based on the $B_{tb+tqb}$ discriminant.  
Variables shown are: leading $b$-tagged jets $p_T$, $W$ boson transverse mass, centrality, defined as the ratio of the scalar sum of the $p_T$ of the jets to the scalar sum of the energy of the jets in the event, and reconstructed $m_t$. 
The presence of the single top quark signal is needed to ensure a good description of the data.

\section{$|V_{tb}|$ measurement}

\begin{figure}
\centering
\includegraphics[width=0.235\textwidth]{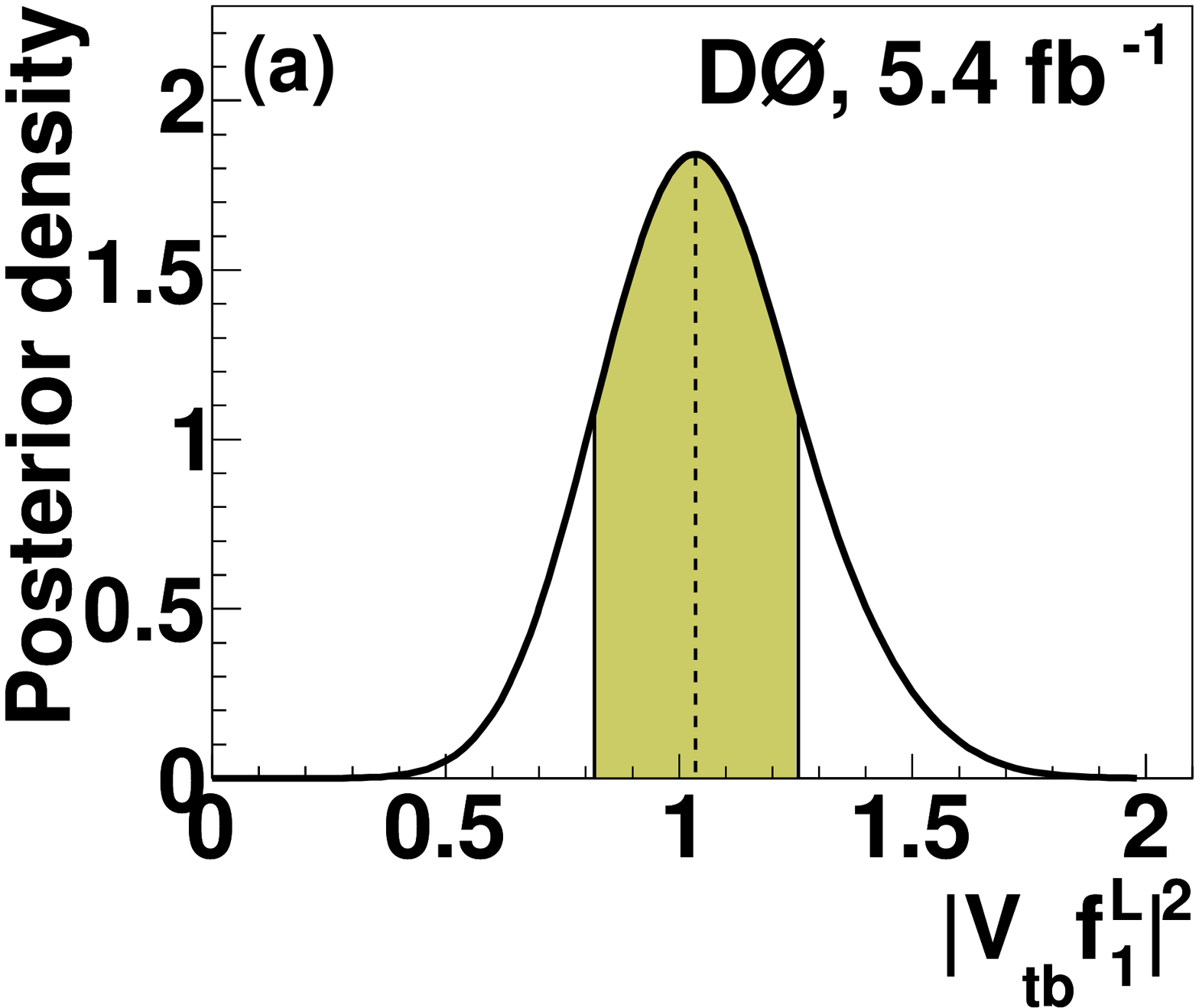}
\includegraphics[width=0.235\textwidth]{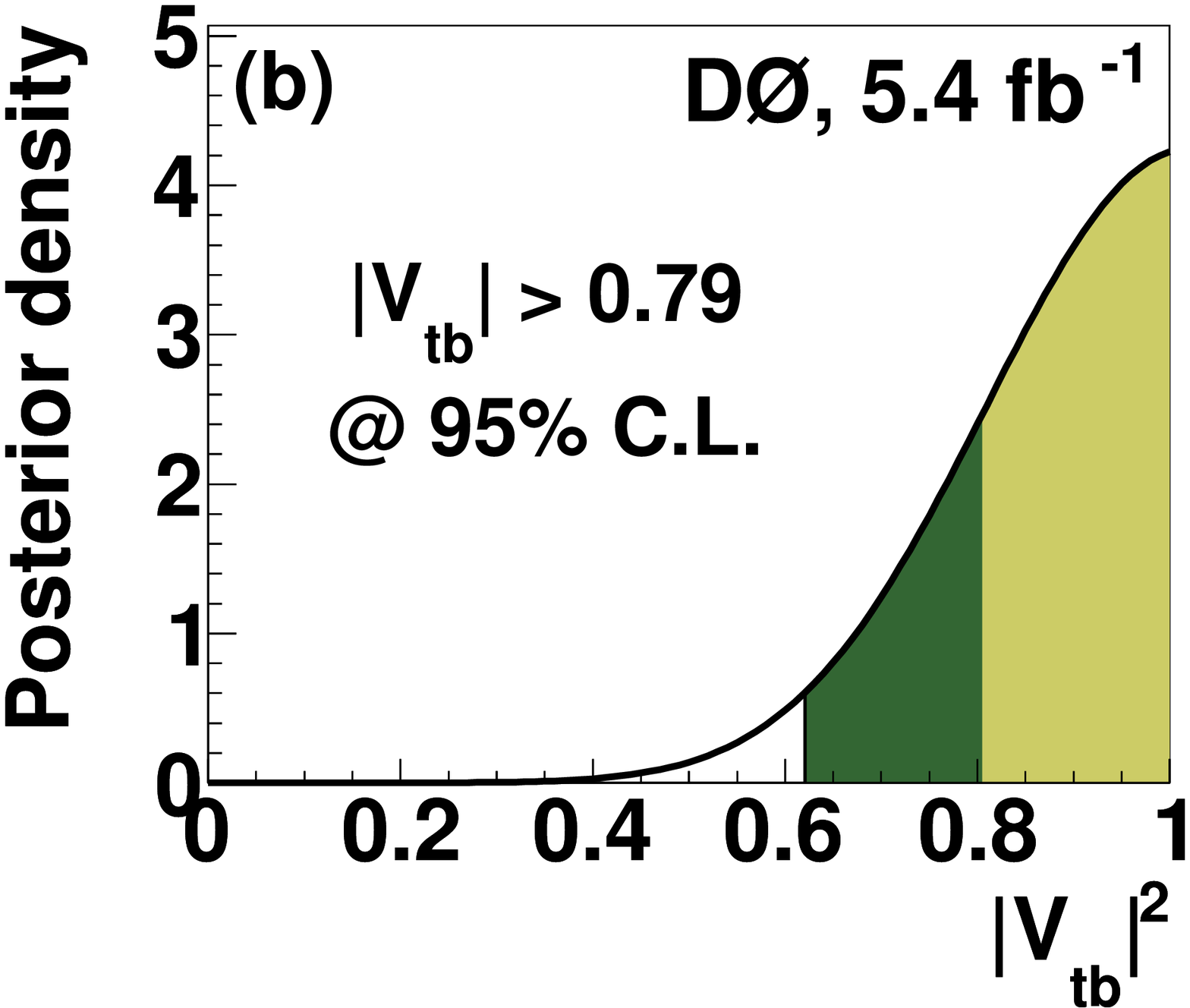}
\caption{The posterior density functions for (a) $|V_{tb}f^L_1|^2$ and (b) $|V_{tb}|^2$. The shaded (dark shaded) band indicates regions of 68\% (95\%) C.L. relative to the peak values.}
\label{fig:vtb}
\end{figure}

The single top quark production cross section is directly proportional to the square of the 
CKM matrix element $|V_{tb}|^2$, enabling 
us to measure $|V_{tb}|$ directly without any assumption on the number of quark families or the unitarity of the CKM matrix~\cite{d0-prd-2008}. 
We assume that SM sources for single top quark production and that top quarks decay exclusively to $Wb$. 
We also assume that the $Wtb$ interaction is CP-conserving and of the $V-A$ type, but maintain the possibility for an anomalous strength 
of the left-handed $Wtb$ coupling ($f_1^L$), which could rescale the single top quark cross section~\cite{flv}. 
Therefore, we are measuring the strength of the $V-A$ coupling, i.e., $|V_{tb}f^L_1|$, which can be $>1$.

We form a Bayesian posterior $|V_{tb}f^L_1|^2$  with a flat prior based on the $B_{tb+tqb}$ discriminant. 
Additional theoretical uncertainties are considered for the $tb$ and $tqb$ cross sections~\cite{singletop-xsec-kidonakis}. 
Using the measured $tb$+$tqb$ cross section, we obtain $|V_{tb}f^L_1| = 1.02^{+0.10}_{-0.11}$. If we restrict the 
prior to the SM region [0,1] and assume $f^L_1=1$, we extract a limit of $|V_{tb}| > 0.79$ at the 95\% C.L. 
Figure~\ref{fig:vtb} shows the posterior density functions for $|V_{tb}f^L_1|^2$ and for $|V_{tb}|^2$, assuming $f^L_1=1$ 
and $0 \leq |V_{tb}|^2 \leq 1$.

\section{Summary}

In summary, we have measured the single top quark production cross section using 5.4~fb$^{-1}$ of data collected by 
the D0 experiment at the Fermilab Tevatron Collider. 
For $m_t=172.5\;\rm GeV$, 
we measure the cross sections for $tb$ and $tqb$ production 
to be 
\[\sigma({\ppbar}{\rargap}tb+X) = 0.68^{+0.38}_{-0.35}\;\rm pb\] 
\[\sigma({\ppbar}{\rargap}tqb+X) =  2.86^{+0.69}_{-0.63}\;\rm pb\] 
assuming, respectively, $tqb$ and $tb$ production rates as predicted 
by the SM. The $tqb$ cross section is consistent with the 
value $\sigma({\ppbar}{\rargap}tqb+X) = 2.90\pm0.59\;\rm pb$ measured in Ref.~\cite{t-channel-new}, 
where we use the same dataset and discriminant but extract the cross section without any 
assumption on the $tb$ production rate.
The total cross section $tb+tqb$ is found to be
\[\sigma({\ppbar}{\rargap}tb+tqb+X) = 3.43^{+0.73}_{-0.74}\;\rm pb\] 
assuming the SM ratio between $tb$ and $tqb$ production. 
All measurements are consistent with the SM predictions for a top quark mass 
of $172.5\;\rm GeV$. 
Finally, we derive a direct limit on the CKM matrix element $|V_{tb}| > 0.79$ at the 95\% C.L. 
assuming a flat prior within $0 \leq |V_{tb}|^2 \leq 1$.

% acknowledgement.tex                             2 August 2011
%
We thank the staffs at Fermilab and collaborating institutions,
and acknowledge support from the
DOE and NSF (USA);
CEA and CNRS/IN2P3 (France);
FASI, Rosatom and RFBR (Russia);
CNPq, FAPERJ, FAPESP and FUNDUNESP (Brazil);
DAE and DST (India);
Colciencias (Colombia);
CONACyT (Mexico);
KRF and KOSEF (Korea);
CONICET and UBACyT (Argentina);
FOM (The Netherlands);
STFC and the Royal Society (United Kingdom);
MSMT and GACR (Czech Republic);
CRC Program and NSERC (Canada);
BMBF and DFG (Germany);
SFI (Ireland);
The Swedish Research Council (Sweden);
and
CAS and CNSF (China).


\begin{thebibliography}{99}
\bibitem{charge-conjugate}
Unless otherwise stated, charge-conjugate states are implied.
\bibitem{eta}
The pseudorapidity is defined as $\eta = -ln[tan(\theta/2)]$,
where $\theta$ is the polar angle relative to the proton beam
direction.
\end{thebibliography}
\end{document}